\documentclass[useAMS,usenatbib]{mn2e}

\usepackage{amssymb}
\usepackage{graphicx}
\usepackage{BibDef}
\usepackage{url}
\def\lsim{\mathrel{\rlap{\lower 3pt \hbox{$\sim$}} \raise 2.0pt \hbox{$<$}}}
\def\gsim{\mathrel{\rlap{\lower 3pt \hbox{$\sim$}} \raise 2.0pt \hbox{$>$}}}

\def\msun{\,\rm{M_\odot}}

\def\ramses{\textsc{RAMSES }}
\def\gizmo{\textsc{GIZMO }}
\title[Super-critical accretion of stellar-mass seeds] {Growing massive black holes through super-critical accretion of stellar-mass seeds}
\author[A. Lupi et al.]{A. Lupi$^{1,3}$, F. Haardt$^{1,3}$, M. Dotti$^{2,3}$, D.Fiacconi$^{4}$, L. Mayer$^{4}$ \& P. Madau$^{4,5}$\\
$^1$DiSAT, Universit\`a degli Studi dell'Insubria, Via Valleggio 11, I-22100 Como, Italy\\
$^2$Dipartimento di Fisica, Universit\`a degli Studi di Milano Bicocca, Piazza della Scienza 3, I-20126 Milano, Italy\\
$^3$INFN, Sezione di Milano-Bicocca, Piazza della Scienza 3, I-20126 Milano, Italy\\
$^4$Center for Theoretical Astrophysics and Cosmology, Institute for Computational Science, University of Zurich,\\Winterthurerstrasse 190, 8057 Zurich, Switzerland\\
$^5$ Department of Astronomy \& Astrophysics, University of California, 1156 High Street, Santa Cruz, CA 95064, USA}
\begin{document}

\date{Draft \today}

\pagerange{\pageref{firstpage}--\pageref{lastpage}} \pubyear{2014}

\maketitle

\label{firstpage}

\begin{abstract}

The rapid assembly of the massive black holes that power the luminous 
quasars observed at $z \sim 6-7$ remains a puzzle. Various direct collapse
models have been proposed to head-start black hole growth from initial
seeds with masses $\sim 10^5$M$_{\odot}$, which can then reach a billion solar 
mass while accreting at the Eddington limit. Here we propose an alternative scenario 
based on radiatively inefficient super-critical accretion of stellar-mass holes 
embedded in the gaseous circum--nuclear discs (CNDs) expected to exist
in the cores of high redshift galaxies. Our sub-pc resolution hydrodynamical 
simulations show that stellar-mass holes orbiting within the central  100 pc of the CND bind to 
very high density gas clumps that arise from the fragmentation of the surrounding gas. Owing
to the large reservoir of dense cold gas available, a stellar-mass black
hole allowed to grow at super-Eddington rates according to the ``slim disc" 
solution can increase its mass by 3 orders of magnitudes within a few million years.
These findings are supported by simulations run with two different hydro codes, \ramses 
based on the Adaptive Mesh Refinement technique and \gizmo based on a new Lagrangian Godunov-type
method, and with similar, but not identical, sub-grid recipes for star formation, supernova feedback, 
black hole accretion and feedback. The low radiative efficiency of super-critical accretion flows are
instrumental to the rapid mass growth of our black holes, as they imply  modest radiative heating 
of the surrounding nuclear environment.
\end{abstract}
\begin{keywords}
black hole - galaxy formation - galaxy evolution.
\end{keywords}

\section{Introduction}
Observations of luminous quasars at very high redshift pose crucial questions on the
formation of massive black holes (MBHs) along the history of the
Universe. The most distant quasar to date, ULAS J1120+0641, lies at
redshift $z=7.084$, and it is believed to be powered by a $\sim 2\times 10^9\msun$ MBH
\citep{mortlock11} that was therefore in place (and shining)
$0.76$ Gyr after the Big Bang. Together with the handful of bright
quasars observed by the {\it Sloan Digital Sky Survey} at $z\gsim 6$
\citep{fan06}, ULAS J1120+0641 sets tight constraints on any model for
the formation and growth of MBHs at early epochs.

It has been long thought that the first MBH seeds were {\it light},
specifically $\sim 100\msun$ remnants of the first generation of
stars, plausibly formed at $z\gsim 20$ \citep[e.g.,][]{madau01,
haiman01,heger03,volonteri03,madau04}. If gas accretion is Eddington-limited, light seeds 
could grow to the supermassive variety by
$z\sim 7$ only if (i) gas accretion continued unimpeded at the Eddington
rate for $\gsim 0.6$ Gyr, and if (ii) the mass-to-light conversion
efficiency of the accretion process was not high, $\epsilon \lsim 0.1$
\citep{tanaka09}. The first condition seems hard to satisfy in
the shallow potential of low--mass dark matter haloes, as radiative
feedback from the progenitor and from BH accretion itself dramatically
affects the gas inflow and its supply to the hole, resulting in
sub--Eddington rates, therefore negligible mass growth \citep[e.g.,][]{wise08,milosavljevic09,alvarez09}. The second condition is problematic too, as it requires
a radiative efficiency well below that proper of accretion onto
rapidly rotating black holes. Indeed, there are mounting
evidences that the most massive holes at high redshifts power radio--loud
AGNs \citep[see, e.g.,][]{ghisellini14}. These are thought to be
associated with Kerr holes -- though observational evidences of the
widely accepted jet--spin connection are, at best, scarce, even in the
well studied Galactic stellar black hole candidates \citep[see, e.g.,][]{russell13}.

Over the last decade, a number of alternatives to the ``light seed
scenario" have been proposed. {\it Heavy} seeds, with masses
as large as $\gsim 10^4\msun$, may form through the direct collapse
of low angular momentum gas at high redshifts \citep[see, e.g.,][]{loeb94,bromm03,koushiappas04,spaans06,mayer10,mayer14}, likely via the intermediate stages of a
supermassive star and a ``quasistar" \citep{begel06,begel08,begel10,dotan11}. For such mechanism to work one needs to, at the same time (i) avoid fragmentation, (ii) 
effectively dissipate angular momentum, and (iii) drive gas towards the centre of the protogalaxies at a rate $\gsim 1\msun$ yr$^{-1}$ \citep[e.g,][]{ferrara13,latif15}. 
There is still no consensus, to date, about where and when such conditions are actually fulfilled. 

In a previous paper \citep[][hereinafter MHD14]{madau14},
we discussed super--critical (i.e., super--Eddington) accretion onto
stellar mass seeds as a possible mechanism for bypassing the above
difficulties. We used the radiatively--inefficient ``slim--disc"
solution \citep{abramowicz88} -- advective, optically thick flows
that generalise the standard Shakura \& Sunyaev solution \citep{shakura73} -- and showed how mildly super--Eddington accretion
significantly eases the problem of assembling MBHs 
in less than a billion year. Because of the
(accretion--rate dependent) low radiative efficiencies of slim discs
around non--rotating {\it as well as} rapidly rotating holes, the
accretion time--scale in this regime is almost independent of the spin
parameter. It is this unique feature of slim discs that makes such
models so appealing.

In MHD14 \citep[see also][]{volonteri15} we briefly
discussed how conditions for super--critical accretion are physically
plausible in the dense environment of high redshift massive
protogalaxies. Here, we elaborate upon this concept by means of high
resolution simulations of a cluster of stellar mass black holes
orbiting the central $\sim 200$ pc of a gas--rich galaxy. We show how
the interplay between gas dynamics and the black holes can easily lead
to the formation of a MBH in the centre of the system within few
million years. Though our simulations are highly idealised and should
be thought as a proof-of-concept of the scenario we are proposing,
they highlight the basic point, i.e. that super--Eddington
accretion in well--formed, evolved galaxies is an attractive route to
the formation of massive black holes. In fact, a population of
stellar mass black holes is expected to reside in the inner $\sim 200$
pc, the circum--nuclear disc can provide enough gas to be accreted, and
negative feedback is negligible in the high--density clumps developed in
the disc.

This is the first of a series of papers devoted at the study of the effect of a radiatively 
inefficient BH feedback on the early growth of stellar mass BHs embedded in a circum--nuclear
gas disc. In a forthcoming paper (Fiacconi et al., in 
preparation) we will discuss the nature and properties of the circum--nuclear gas 
by means of a fully cosmological, high--resolution simulation of the formation of a gas--rich massive disc
galaxy at $z > 6$.

\section{Simulations}

We perform a suite of 6 simulations, using the adaptive mesh
refinement (AMR) code \ramses \citep{teyssier02} (``\_R''
runs) and the new mesh--free code \gizmo \citep{hopkins14}
(``\_G'' runs), which implements a new Lagrangian method
to solve the hydrodynamic equations (similar to Godunov--type schemes). 
The two codes are equipped with  similar, but not identical sub-grid
recipes for star formation, supernova feedback, BH accretion and BH feedback.
While RAMSES is a well-tested AMR code which has been already successfully
employed to study circum--nuclear gas discs in merger remnants \citep{chapon13}, GIZMO employs a novel numerical technique which is in principle
well suited to study the flow in a highly dynamical situation while conserving 
angular momentum and limiting numerical diffusion, with advection errors that are
smaller than grid-based cartesian AMR codes such as RAMSES \citep{hopkins14}. GIZMO 
is thus supposed to combine the strengths of grid-based and SPH codes while 
retaining the tree-based gravity solver inherited from GADGET3, which
guarantees high accuracy as well as fast computation for self-gravitating
discs like those under study here. The use of two different powerful
numerical techniques is aimed at checking the robustness and reproducibility of our results. 

\subsection{Initial conditions}

\begin{figure}
\centering
\includegraphics[scale=0.45]{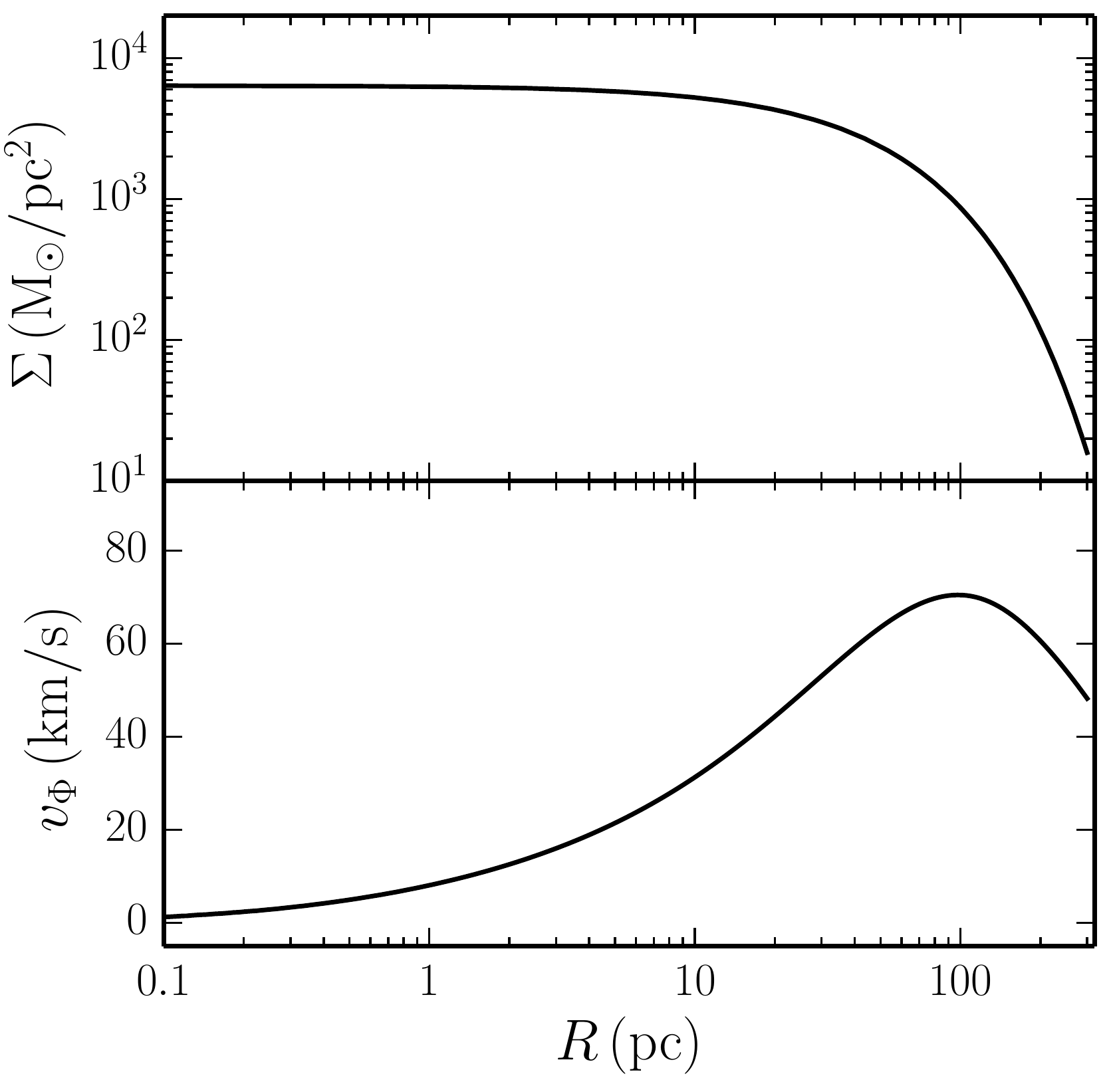}
\caption{\rm{Initial surface density (top panel) and circular velocity (bottom panel) profiles 
of the circum--nuclear disc.}}
\label{fig:cvel}
\end{figure}

In our simulations we consider the nuclear region of a high--redshift
massive spiral galaxy, where we assume a gaseous circum--nuclear disc
(CND hereafter) is hosted. We model the CND following an
exponential surface density profile (see Fig.~\ref{fig:cvel}, top
panel), with total mass $10^8\msun$ and scale radius 50 pc. The disc
is embedded in a stellar spherical background following an Hernquist
profile, with scale radius 100 pc and total mass $2\times
10^8\msun$. Note that the adopted gas mass from the CND is at the lower end
of that of CNDs observed in low z merger remnants \citep[e.g.,][]{medling14}, and
should thus be regarded as a conservative choice for the purpose of our paper.
At high redshift galaxies are indeed expected to be more gas rich, which is supported
by both theoretical arguments and observational evidence.

\begin{figure}
\includegraphics[scale=0.43]{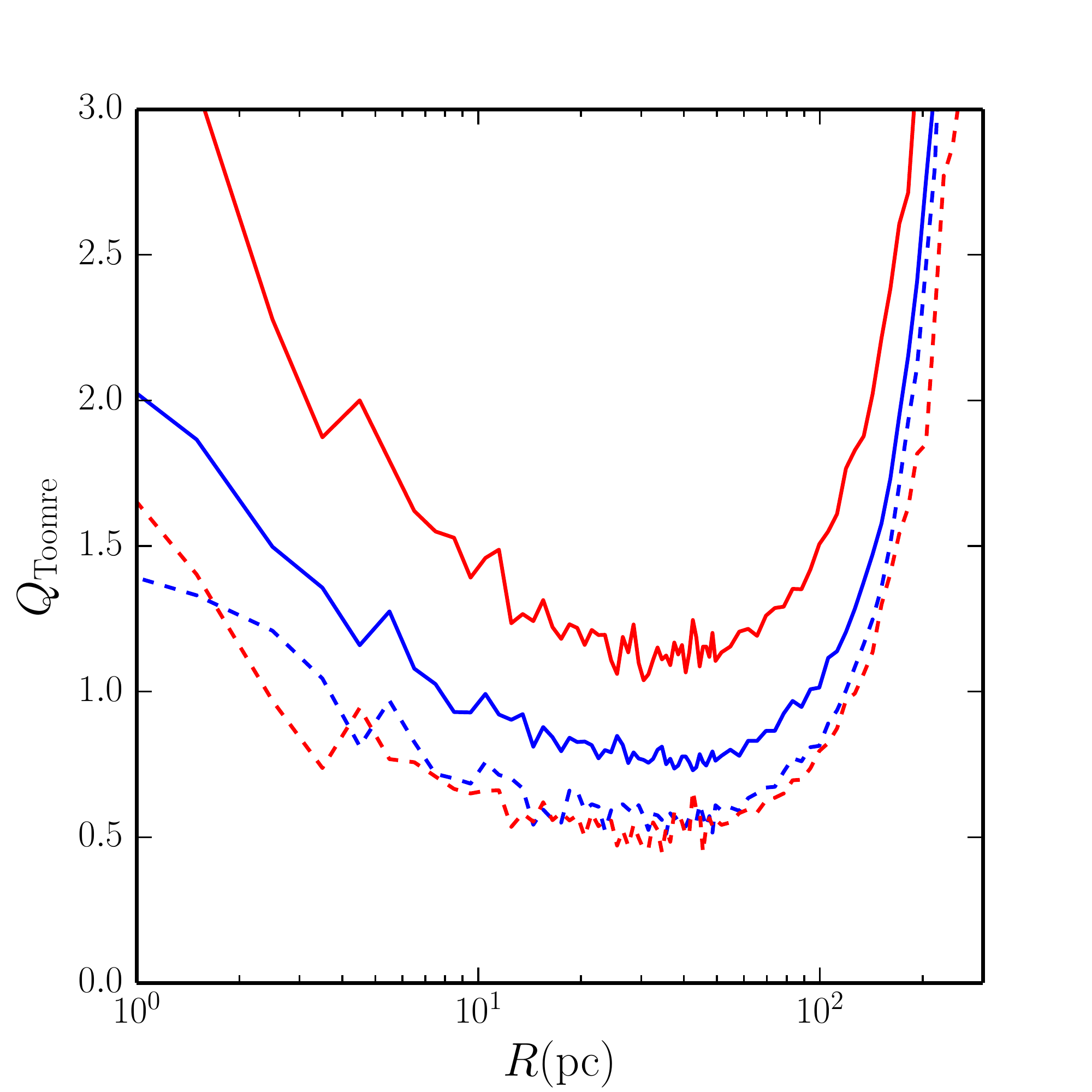}
\caption{\rm{Toomre parameter for the isolated CNDs at $t=0$ Myr (solid lines) and $t=0.05$ Myr (dashed lines). The blue line corresponds to an initial temperature $T=10^4$ K and the red one to an initial temperature $T=3\times 10^4$ K.}}
\label{fig:toomre}
\end{figure}

The disc, modelled as an ideal gas with polytropic index
$\gamma=5/3$, is set in hydrostatic equilibrium in the
global potential with an initial temperature $T=10^4$ K. The initial conditions are produced using the
public code \textsc{ GD\_BASIC}, described in \citet{lupi15a}. We
show in Fig.~\ref{fig:cvel}, bottom panel, the velocity profile of
the gas component in the disc. 
The CND is Toomre unstable since the beginning of the run in the region between 10 pc and 70 pc, as reported in Figure~\ref{fig:toomre} (blue solid line). In order to check if such low temperature could result in  overestimated clumpiness we perform two runs of the disc in isolation, one starting with the above mentioned temperature and one with $T=3\times 10^4$ K (the equilibrium temperature in the case of a disc heated by UV background radiation at $z\sim 8$). We do not include UV background radiation in the simulation, assuming the disc will become self--shielded. Despite the initially higher temperature, we find that after only few $10^4$ years the disc temperature drops to $7\times 10^3$ K in both runs, converging to an identical profile for the Toomre parameter $Q$ (Figure~\ref{fig:toomre}, dashed lines).

We then assume that previous star-formation episodes left a population of stellar mass black holes in the galaxy nucleus. The mass of such ``black hole seeds" $M_{\rm BH}$ is alternatively set 20 or 100M$_\odot$. We initially distribute the BHs uniformly within the inner 150 pc of the CND.
The BHs lay in the disc plane and have an initial velocity equal to the local circular velocity. We add a randomly oriented 
velocity component sampled from a normal distribution with standard deviation $\sigma\sim 20$\% of the maximum circular velocity.

\subsection{\ramses Eulerian simulations}

We perform two simulations with RAMSES at two different spatial resolutions, namely 0.4 pc (``low'' runs) and 0.1 pc (``med'' runs).  The mass resolution is $10^3 \msun$ at the
quasi--Lagrangian threshold for refinement. We include the radiative
cooling of the gas adopting the standard prescriptions employed 
in the code \citep[see][for details]{teyssier13}. In order to prevent spurious fragmentation at the highest
refinement level we add a polytropic pressure term to the gas
component (described as a polytrope with $\gamma=5/3$ and temperature
$10^3$ K at $2\times 10^5\rm\, cm^{-3}$), ensuring to resolve
the Jeans length with at least 4 cells at the highest refinement
level.

We set a star formation density and temperature threshold of
$\rho_{\rm thr} = 2\times 10^5$ cm$^{-3}$ and $T_{\rm thr} =2  \times
10^4$ K, and a typical star formation time--scale of 1 Myr. We also assume a
time delay between star formation and the corresponding SNa explosion event of 1
Myr, with a SNa yield of 0.15 (corresponding to stars with masses
above $8\msun$ for a Salpeter IMF). In order to model non thermal
processes associated with SNa events, we include the blast--wave like
feedback described in \citet{teyssier13}. In this feedback recipe the
SNa energy budget is decoupled from the thermal energy of the gas, preventing, for a typical
timescale $\simeq 20$ Myr, the gas to radiatively cool. In our runs we assume a primordial gas
composition and we include subsequent metal pollution due
to SNae.

%\footnote{As the system evolves the clump $M_{\rm top}$ binds to grows in mass, reaching a cell mass peak of $\sim 700\msun$ after 2 Myr. However, at this time, $M_{\rm top}$ is already more massive than $10^3\msun$}. 

\subsection{\gizmo Lagrangian simulations}

Since the public version of \gizmo only includes basic
hydrodynamics and gravity, we implement in the code the additional
recipes necessary to model gas cooling, star formation, type II
SNa feedback and gas accretion onto BHs, in a fashion similar to \ramses implementations.

We include radiative cooling by means of the
\textsc{GRACKLE}\footnote{\url{http://grackle.readthedocs.org}}
chemistry and cooling library, which provides both equilibrium and non--equilibrium chemistry \citep{bryan14,kim14}. 
In our runs we employ the equilibrium cooling curve for primordial species
(atomic \textsc{H} and \textsc{He}), and tabulated metal cooling and
heating from \textsc{Cloudy} \citep{ferland13}.

Gas particles are eligible for star formation when they match the same criteria for density and temperature adopted in our \ramses runs and 
belong to a converging flow (i.e., $\nabla\cdot \mathbf{v}<0$). 
Resulting star particles are generated locally according to the
Schmidt law \citep{kennicutt98}, and using a stochastic prescription as described in
\citet{stinson06}.  We model SNa feedback assuming that after
a typical timescale of 1 Myr stars above $8\msun$ explode as type II
SNae, releasing $10^{50}\rm\, erg/\msun$ in the form of purely thermal energy. 

The non thermal processes that energise the SNa blast--wave 
have a typical dissipation timescale much longer than that of the thermal energy. 
Therefore, we implement a ``delayed cooling'' prescription as follows: 
we temporarily inhibit radiative cooling for gas particles within
the SNa maximum extension radius $R_{\rm E}$ defined as in \citet{chevalier74}.
%$R_{\rm E}=10^{1.74}
%E_{51}^{0.32}n_0^{-0.16} \tilde{P}_{4}}^{-0.20}$ pc, where $E_{\rm
%  SN}=E_{51}10^{51}$ erg, $n_0$ is the ambient hydrogen density and
%$\tilde{P}_{4}}=10^{-4}P_0 k_{\rm B}^{-1}$ with $P_0$ the ambient
%pressure \citep[see][for details]{chevalier74}. 
The SNa energy is then distributed among the gas particles lying within $R_{\rm E}$, according to their distance from the SNa, weighted through the kernel function used in the code. This is different from
the approach taken in the \ramses runs, in which the energy is wholly released 
within the cell hosting the progenitor. 
In \gizmo runs we limit the cooling delay time to 5 Myr only (i.e., 4 times smaller than what assumed in our \ramses runs). We checked that this set up provides consistent results between \ramses and \gizmo feedback implementations.

We perform three simulations allowing for two different gravitational
resolutions, i.e., 0.16 pc (``low'' runs) and 0.02 pc (``high'' run). We set the same gravitational resolution for gas particles and BH particles. We use $10^5$ particles for the ``low'' runs and
$10^7$ for the ``high'' run, corresponding to a mass resolution of
$10^3\msun$ and $10\msun$, respectively.  
In our runs we use the finite mass mode available in the code, in which mass transfer between particles is forbidden, so that our simulations are purely Lagrangian.

\subsection{BH accretion}

In RAMSES runs we evaluate the accretion onto the stellar mass BHs using the so-called  ``flux accretion'' prescription. In such a scheme the accretion rate is the
mass flux rate within the BH accretion zone, which consists of a sphere of radius equal to 4 cells, i.e., 
$\dot{M}_{\rm acc}=\int-\nabla\cdot [\rho\Delta \mathbf{v}]
d^3\mathbf{x}$, where the integral is over the volume of the accretion zone and $\Delta \mathbf{v}$ is
the gas--BH relative velocity \citep[see][for a detailed description of the implementation]{bleuler14}.

In order to get a more accurate BH dynamics and to best resolve the
accretion rate, we force the region near to each BH to
always be at the maximum refinement level, as described in
\citet{lupi15a}. Forcing
the resolution close to the BHs at the highest possible level guarantees that 
nearby cells own a mass $\lsim 5\times M_{\rm BH}$ during the whole BH accretion history. This allows us to set 
$20\msun$ as the initial mass of the BHs in our \ramses runs.\\

\noindent
In GIZMO runs we allow gas accretion onto the stellar--mass BHs for particles lying 
within a distance from the hole than encompasses an
effective number of gas particles $N=32$. Such {\it kernel radius} $h$ is therefore 
implicitly defined through the following relation:
\begin{equation}
\frac{4}{3}\pi h^3\sum_j W(z_j)=32.
\end{equation}
The {\it volume partition kernel} $W(z_j)$ is a function of the normalised distance to the BH 
$z_j\equiv |{\mathbf x}_j-{\mathbf x}_{\rm BH}|/h$, where ${\mathbf x}_j$ and ${\mathbf x}_{\rm BH}$ are the position 
vectors of the $j$-th gas particle and of the accreting BH, respectively \citep[for further details see][]{hopkins14}.  
Since the kernel radius strongly depends upon the gas density, our scheme would
typically overestimate the accretion rate when a BH moves in a low
density medium. In order to overcome this problem we enable accretion onto a BH 
only when the kernel radius is $\leq 10$ times the softening length of the sink particle.
%$h<10\varepsilon_{\rm BH}$, where $\varepsilon_{\rm BH}$ is the BH softening. 
The accretion rate is then computed as described for \ramses runs. 

%In our \gizmo runs we set all the sub-grid physics parameters
%like in \ramses.
Because of the large
mass of gas particles the dynamics of BHs as light as 20 $\msun$ cannot be properly solved for. 
Therefore, we start from a larger
initial BH mass, i.e., $M_{\rm BH}= 100\, \msun$. With such a choice, BHs in the ``high'' runs have resolved
dynamics since the very beginning of the simulation. In the ``low'' case,
the initial BH dynamics and growth is instead affected by the lack of
mass resolution. However,  as will be discussed in the next section, some BHs grow above
$1000\, \msun$ in a very short time, making dynamics quickly reliable.

With respect to the standard Bondi-Hoyle model, this recipe does not make any geometrical assumption for the gas flow, allowing for a more accurate estimation of the accretion rate, where the effect of angular momentum on the resolved scales is taken into account. However, despite the high resolution reached with the ``high\_G" run, we are unable to properly follow the gas from sub--parsec scales down to the accretion disc scale. This resolution limit could result in a overestimated and more efficient accretion. Such a convergence study is beyond the scope of the present work.

\subsection{BH feedback}

The public release of \ramses already provides a prescription for AGN thermal feedback \citep{dubois14a}, in which the radiation produced by accretion is stored until the total budget is large enough to heat up the surrounding gas to at least $10^7$ K. This prescription has been initially proposed by \citet{booth09} to prevent the gas from immediately loose the small amount of additional thermal energy gained after each time--step, which would result in an ineffective feedback. The \ramses prescription assumes a fixed accretion radiative efficiency $\epsilon=0.1$, and a fixed fraction $=15\%$ of the accretion energy to be released to gas.
In our simulations we suitably modify this recipe to include the effects of accretion in the fashion of slim disc \citep{sadowski14}. To this aim, we estimate $\epsilon$ using the analytical fit to the numerical results by \citet{sadowski14} 
provided by \citet{madau14}:
\begin{equation}
\epsilon=\frac{r}{16}A(a)\left[\frac{0.985}{r+B(a)}+\frac{0.015}{r+C(a)}\right],
\end{equation}
where $r=\dot{M}_{\rm E}/\dot{M}$. Here $\dot{M}_{\rm E}=16L_{\rm E}/c^2$ where $L_{\rm E}$ 
is the Eddington luminosity. $A,B,C$ are fitting functions scaling with the BH spin $a$ as
\begin{eqnarray}
A(a)&=&(0.9663  - 0.9292a)^{-0.5639},\\
B(a)&=&(4.627  - 4.445a)^{-0.5524},\\
C(a)&=&(827.3  - 718.1a)^{-0.7060}.
\end{eqnarray}

At each accretion event we compute the released energy allowed to feedback on nearby particles using this new value for $\epsilon$ instead of the fixed value 0.1, while the spin is always fixed at $a=0.99$ for all BHs.  
We implement in \gizmo the same prescription for BH feedback. In all simulations we do not include
other possible forms of BH feedback, e.g., momentum-driven.

Finally, in order to check whether super--critical
accretion is instrumental in leading to very large $M_{\rm BH}$ in a 
short time, we perform two \gizmo runs setting the radiative efficiency to its custom value, $\epsilon=0.1$ (low\_G\_0.1 and high\_G\_0.1 runs).
We report the details of our six simulations in
table~\ref{tab:suite}.
\begin{table*}
\centering
\begin{tabular}{lccccc}
Run & Resolution & BH mass  & Accretion radius & $\epsilon$\\
      &  (pc)	& $(\msun)$ &	(pc) &  \\
\hline
\hline
low\_R & 0.40 & 20 & 1.6 & Slim\\
med\_R & 0.10 & 20 & 0.4 & Slim\\
\hline
low\_G & 0.16 & 100  & $<1.6$ & Slim\\
high\_G & 0.02 & 100 & $<0.2$ & Slim\\
\hline
low\_G\_0.1 & 0.16 & 100 & $<1.6$ & 0.1\\
high\_G\_0.1 & 0.02 & 100 & $<0.2$ & 0.1\\
\hline
\hline
\end{tabular}
\caption{\rm{Settings of our simulation suite. The second column
    reports the gravitational resolution (for \_G runs) and the
    highest refinement level resolution (for \_R runs). The fourth
    column is the accretion radius, which is fixed to 4
    cells for \_R runs and depends on the smoothing length for \_G
    runs. The last column indicates the type of accretion recipe used.}}
\label{tab:suite}
\end{table*}

\section{Results}\label{sec:results}

\begin{figure*}
  \includegraphics[width=0.3\textwidth]{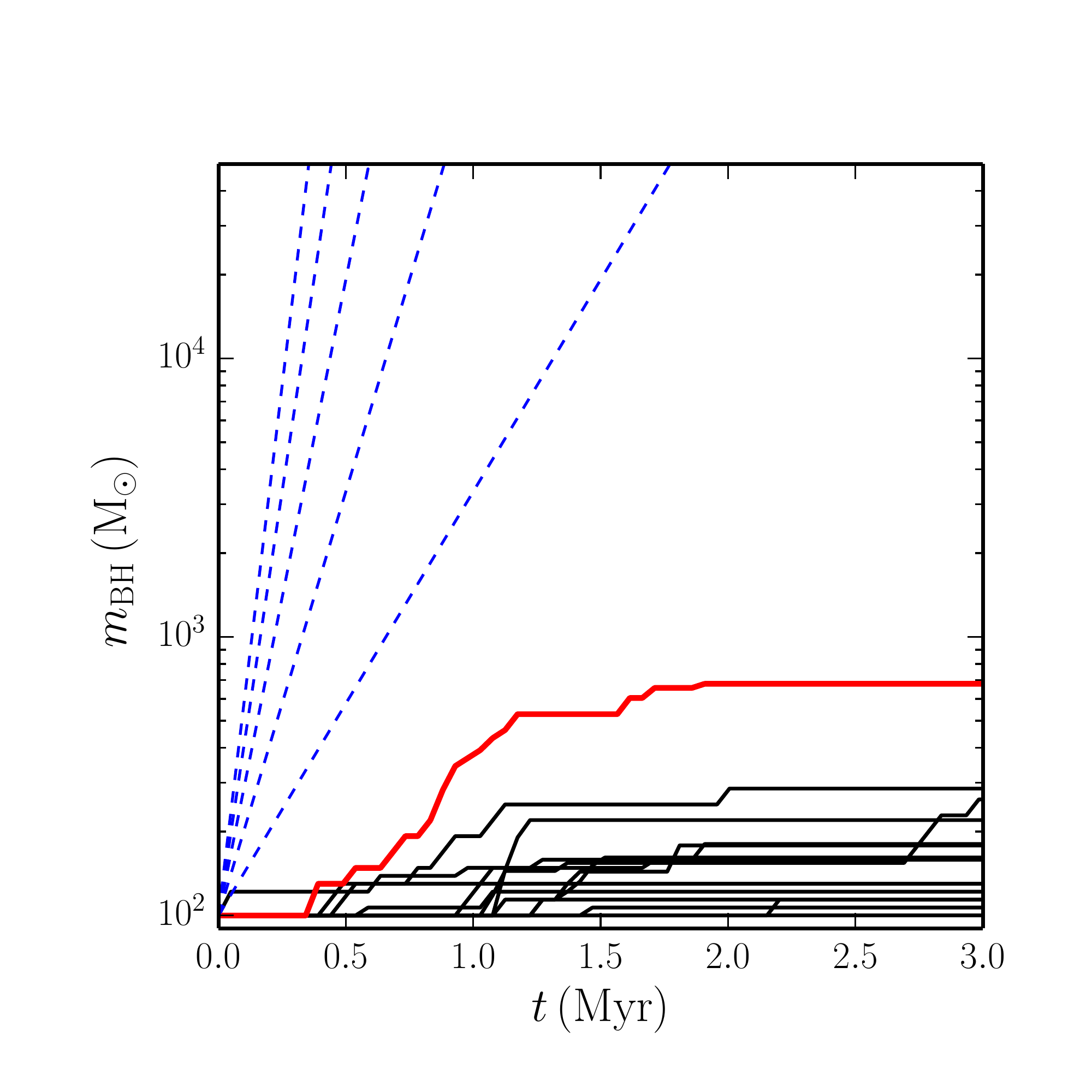}
  \includegraphics[width=0.3\textwidth]{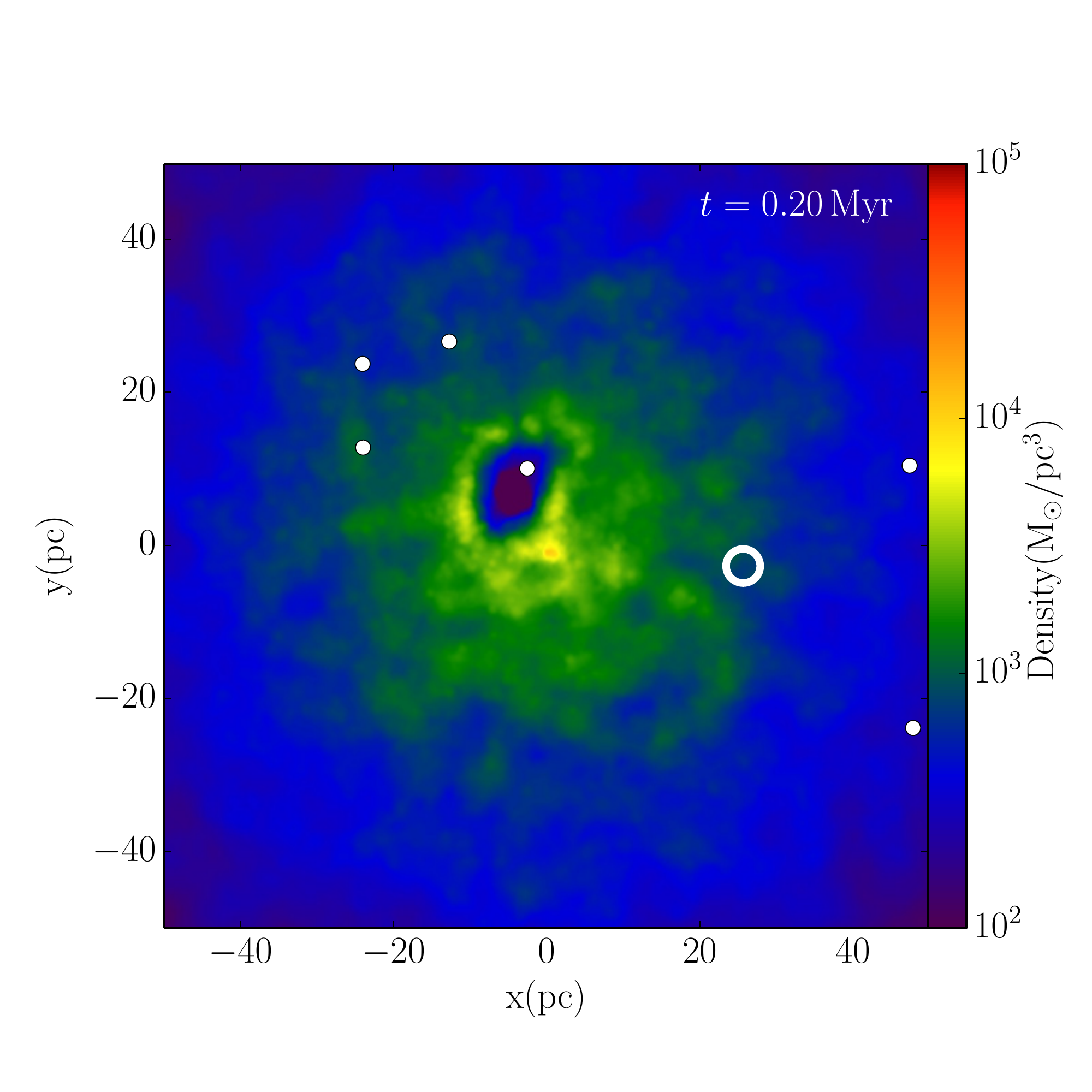}
  \includegraphics[width=0.3\textwidth]{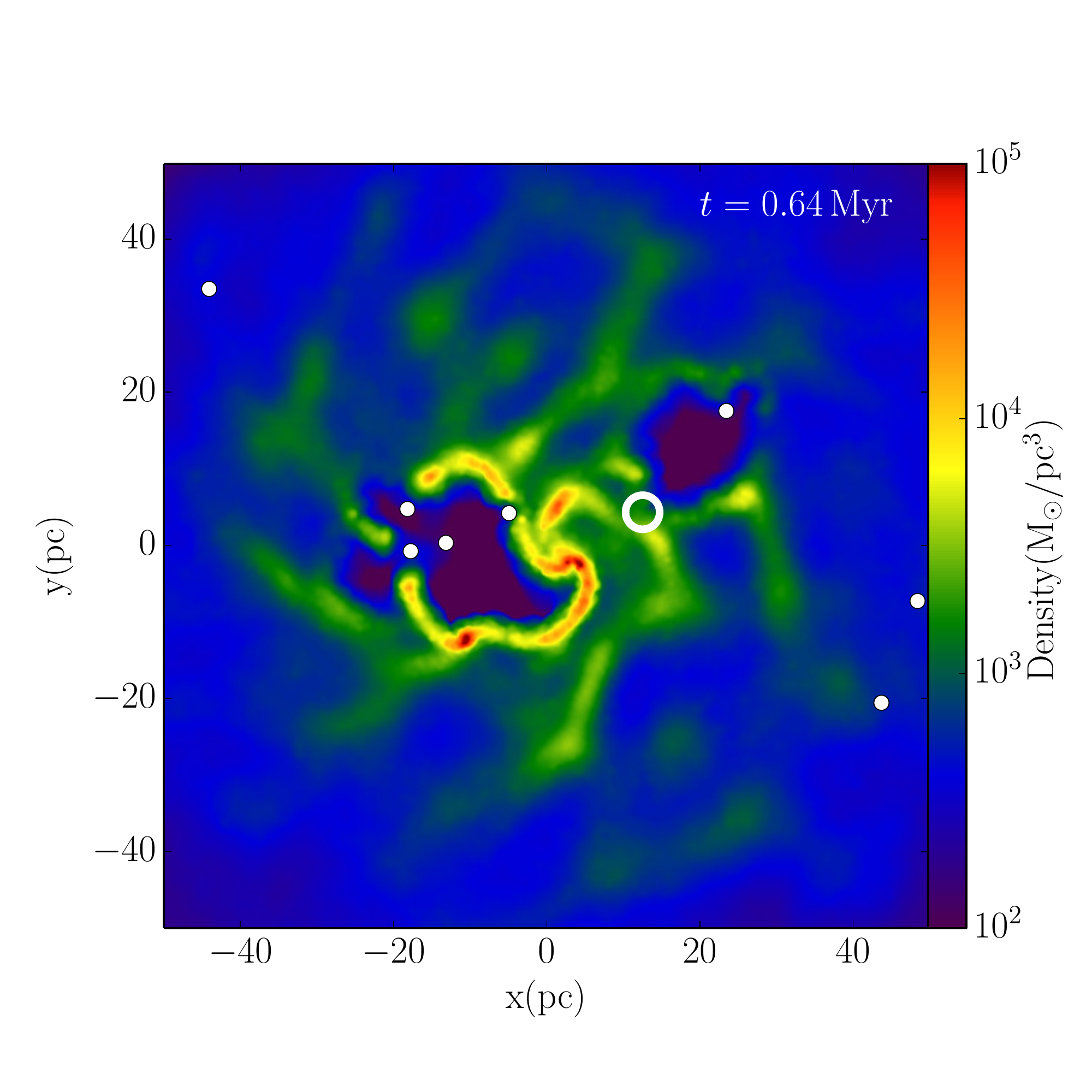}\\
  \includegraphics[width=0.3\textwidth]{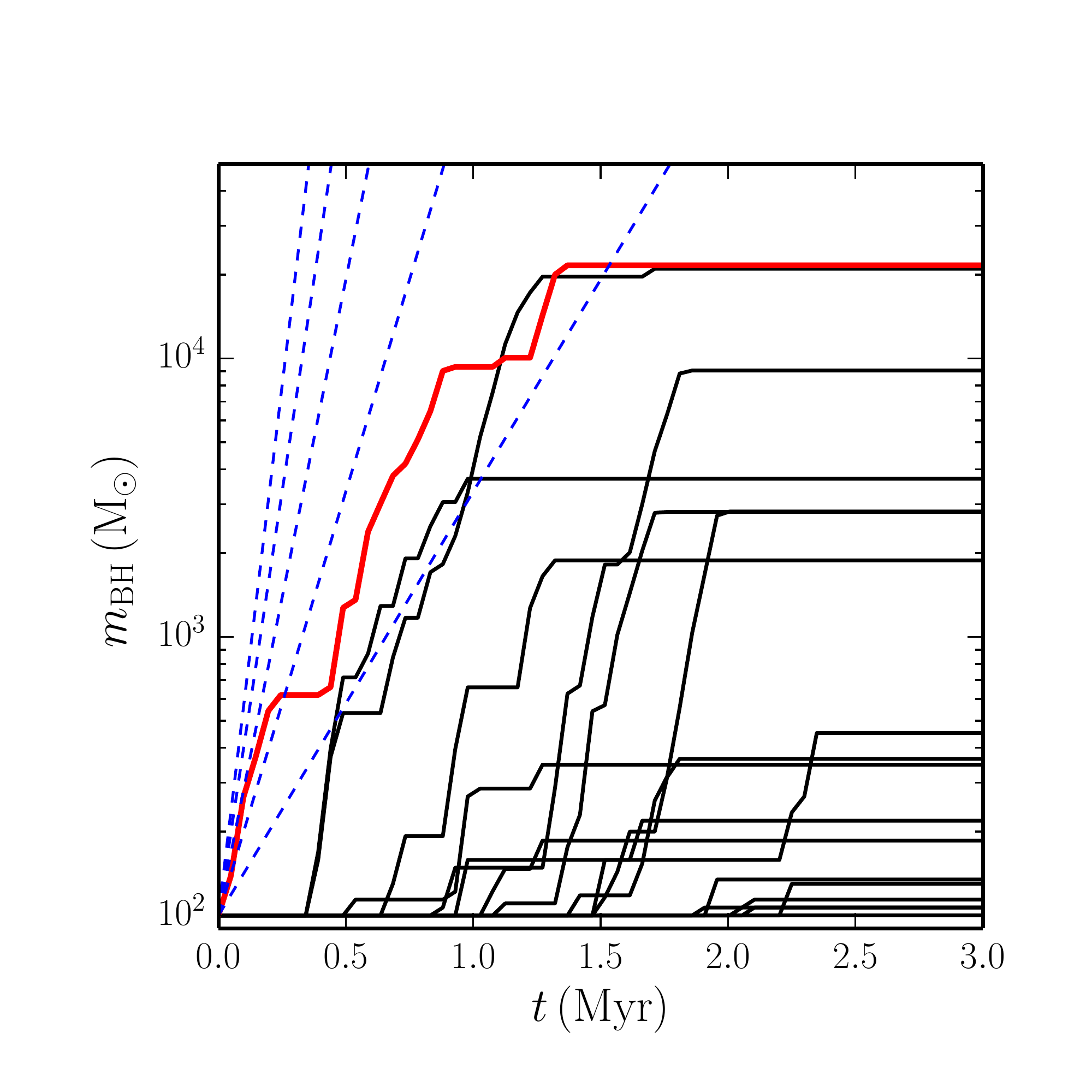}
   \includegraphics[width=0.3\textwidth]{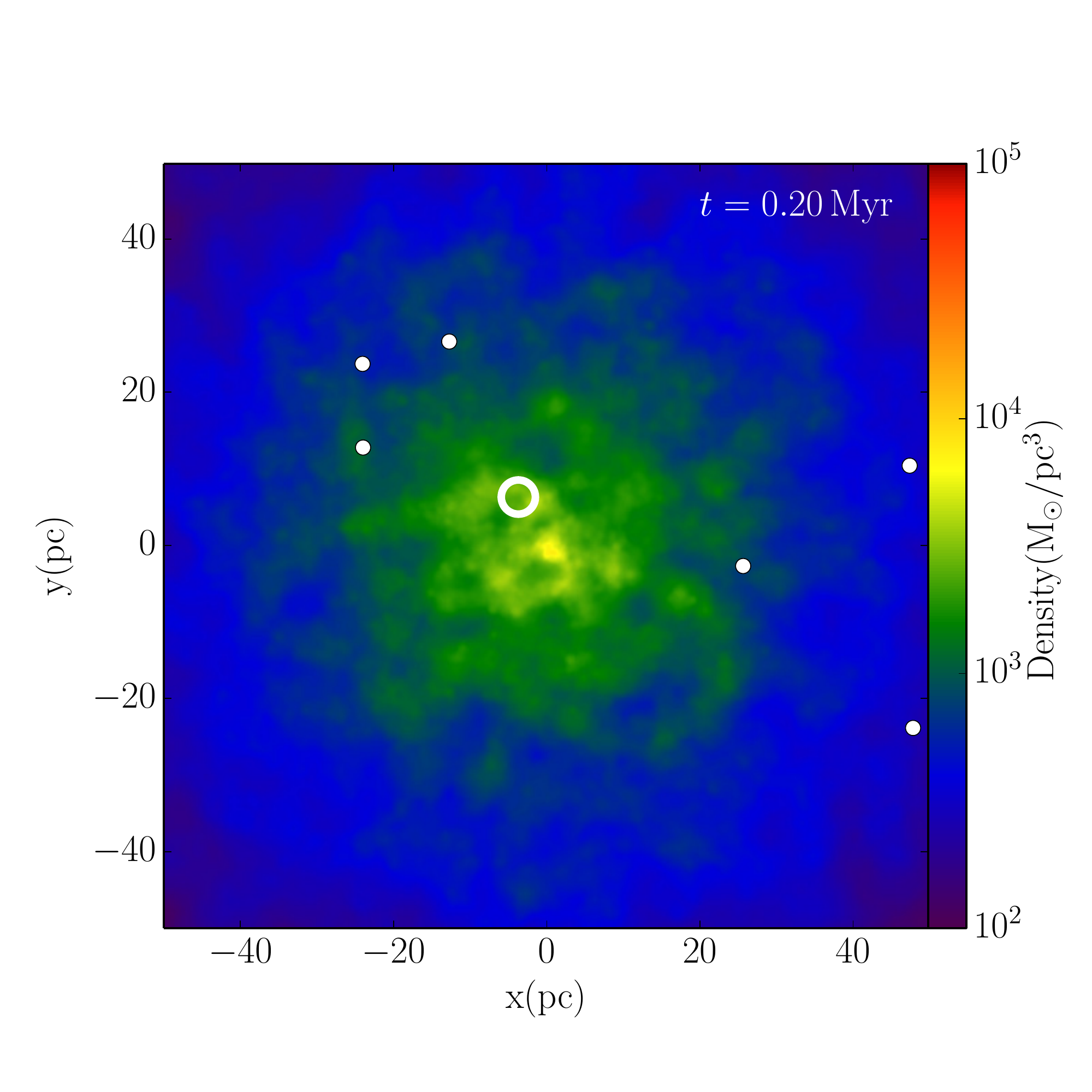}
  \includegraphics[width=0.3\textwidth]{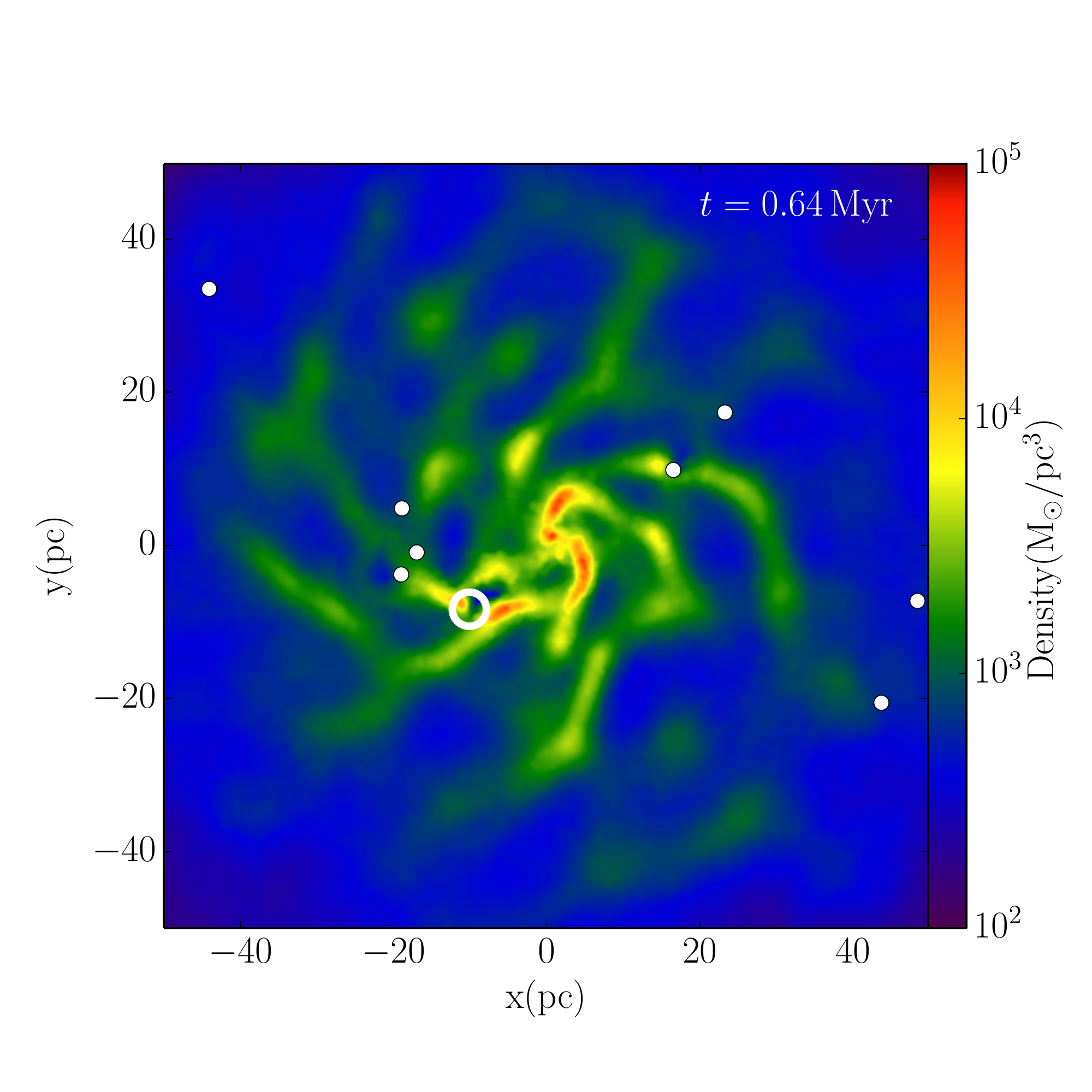}
\caption{\rm{Left panels: BH masses as a function of time for runs
    low\_G\_0.1 (top panel) and low\_G (bottom panel). The red
    lines correspond to the most massive BHs (BH$_{\rm top}$) at the
    end of the runs, while the blue dashed lines trace accretion
    histories at fixed Eddington ratios of 500, 400, 300, 200 and 100,
    respectively. Central and right panels: gas density maps for the
    two runs at $t=0.2$ and 0.64 Myr, respectively. The white dots mark the
    positions of the BHs.}}
\label{fig:mtime_low}
\end{figure*}

Figure~\ref{fig:mtime_low} shows the comparison between the low
resolution \gizmo runs with (low\_G) and without
(low\_G\_0.1) the slim disc implementation. All the other
simulation parameters are the same in the two simulations. 
It is immediately clear from the comparison that whenever a BH undergoes an intense
accretion episode, the large feedback energy available in the
radiatively efficient low\_G\_0.1 case evacuates the BH
surroundings, efficiently limiting further BH growth. In the low\_G
case, on the contrary, even accretion rates significantly higher than
$\dot{M}_{\rm E}$ result in moderate luminosities that do not impact on 
the densest gas clumps, and therefore BHs can grow considerably faster. As an
example, in the low\_G run the most mass growing BH (that will be
referred to as BH$_{\rm top}$ in all runs hereafter, red line in
the bottom--left panel of figure~\ref{fig:mtime_low}) reaches a mass larger
by up to 2 order of magnitudes compared to the corresponding BH$_{\rm top}$ in the
radiative efficient case at the end of the simulation (red line in the top--left panel panel). 
The low radiative efficiency of slim discs has then a double effect: first, for any given accretion rate BHs grow 
faster simply because less mass is lost as radiation (the ``$(1-\epsilon)$-effect"); second, the reduced radiative efficiency results in a reduced feedback on the accreting gas, and larger accretion 
rates are therefore possible (the ``$\dot M$-effect"). 

\begin{figure*}
  \includegraphics[width=0.3\textwidth]{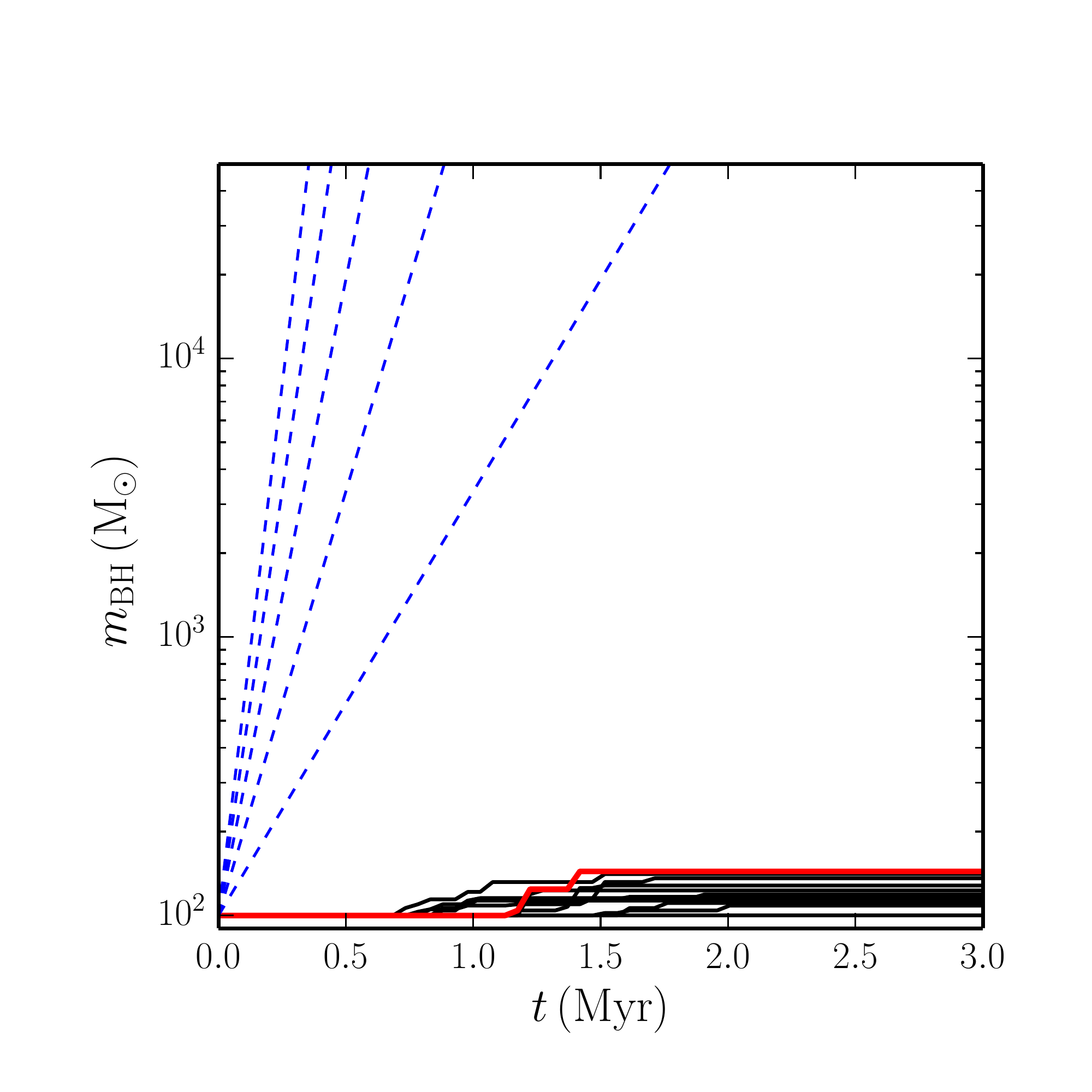}
  \includegraphics[width=0.3\textwidth]{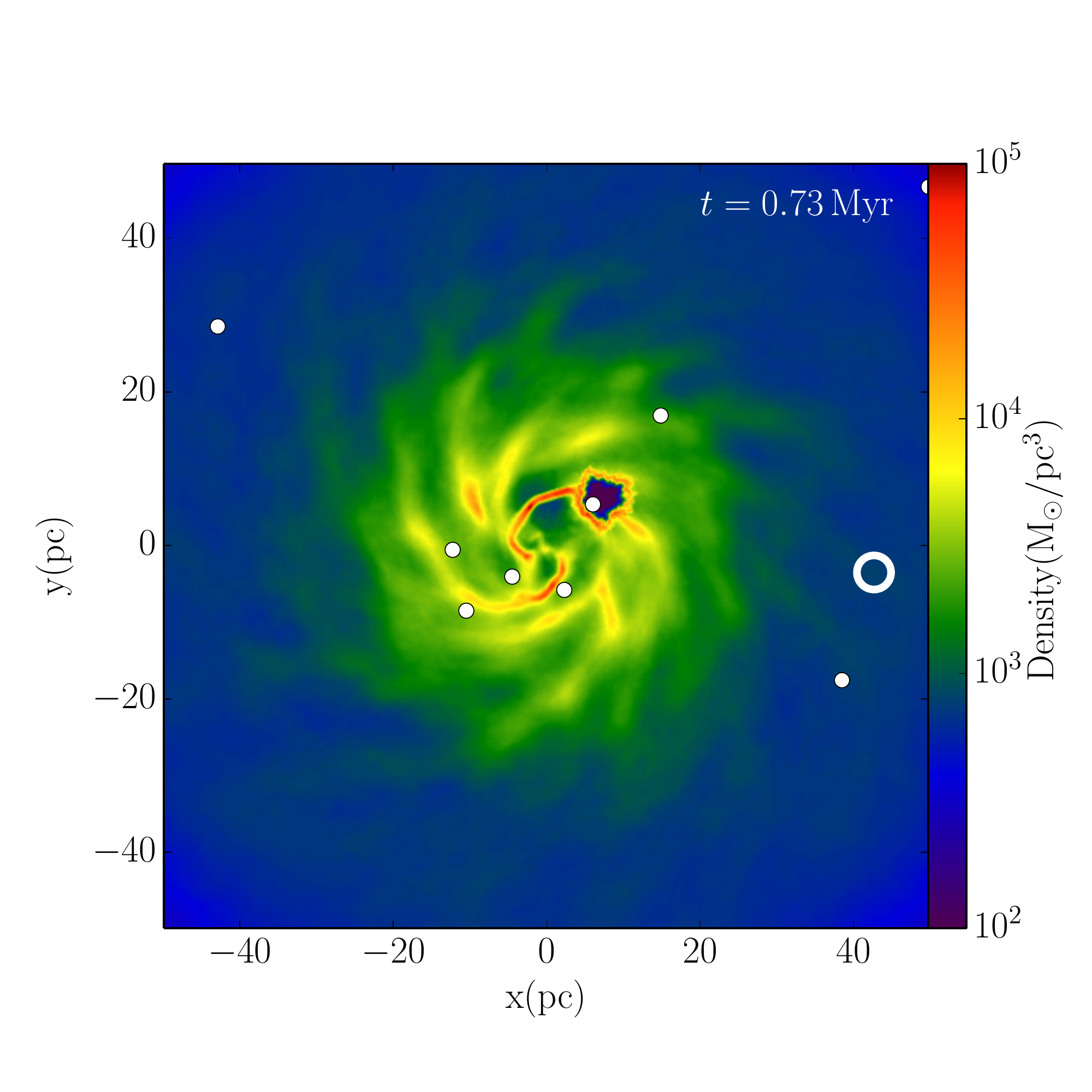}
  \includegraphics[width=0.3\textwidth]{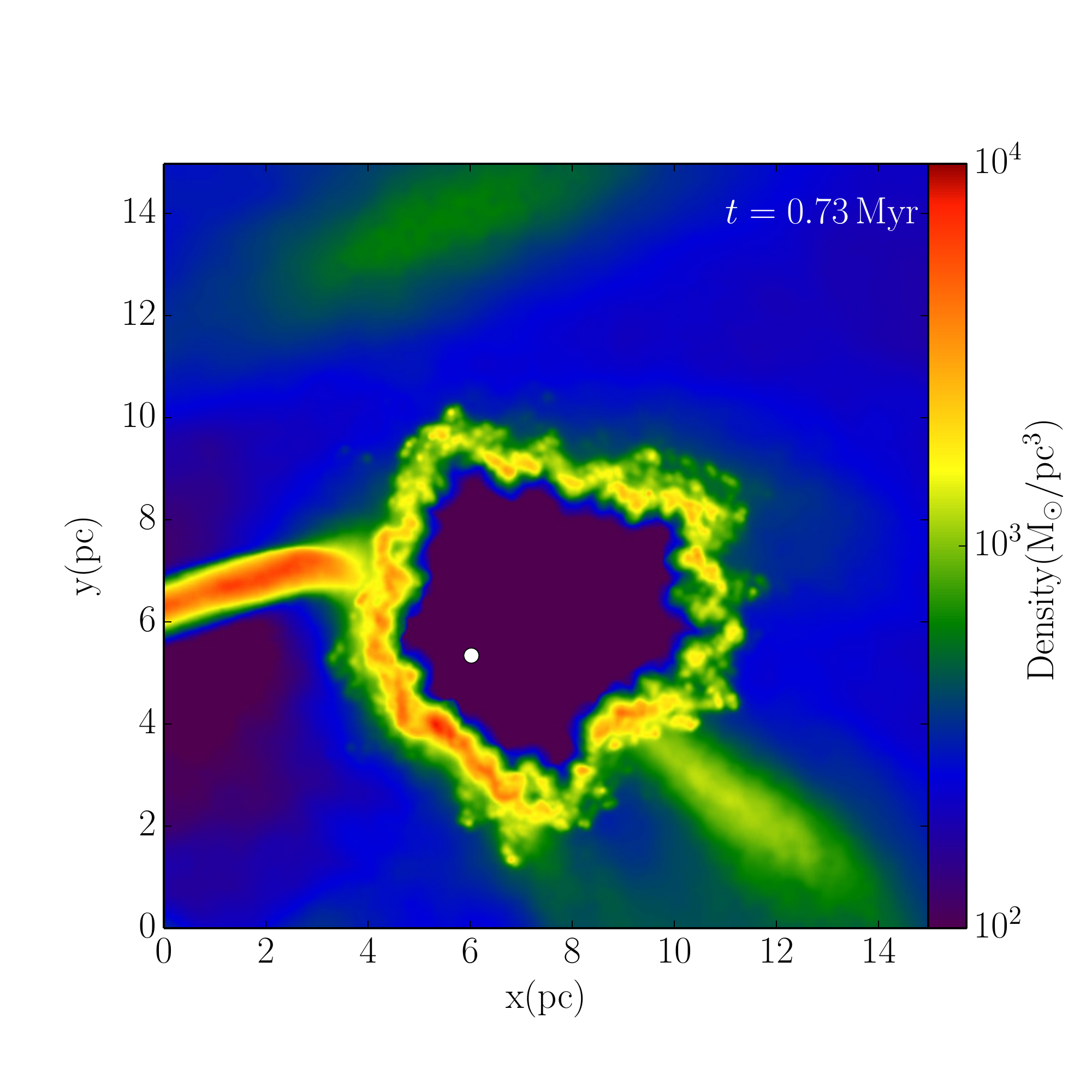}\\
  \includegraphics[width=0.3\textwidth]{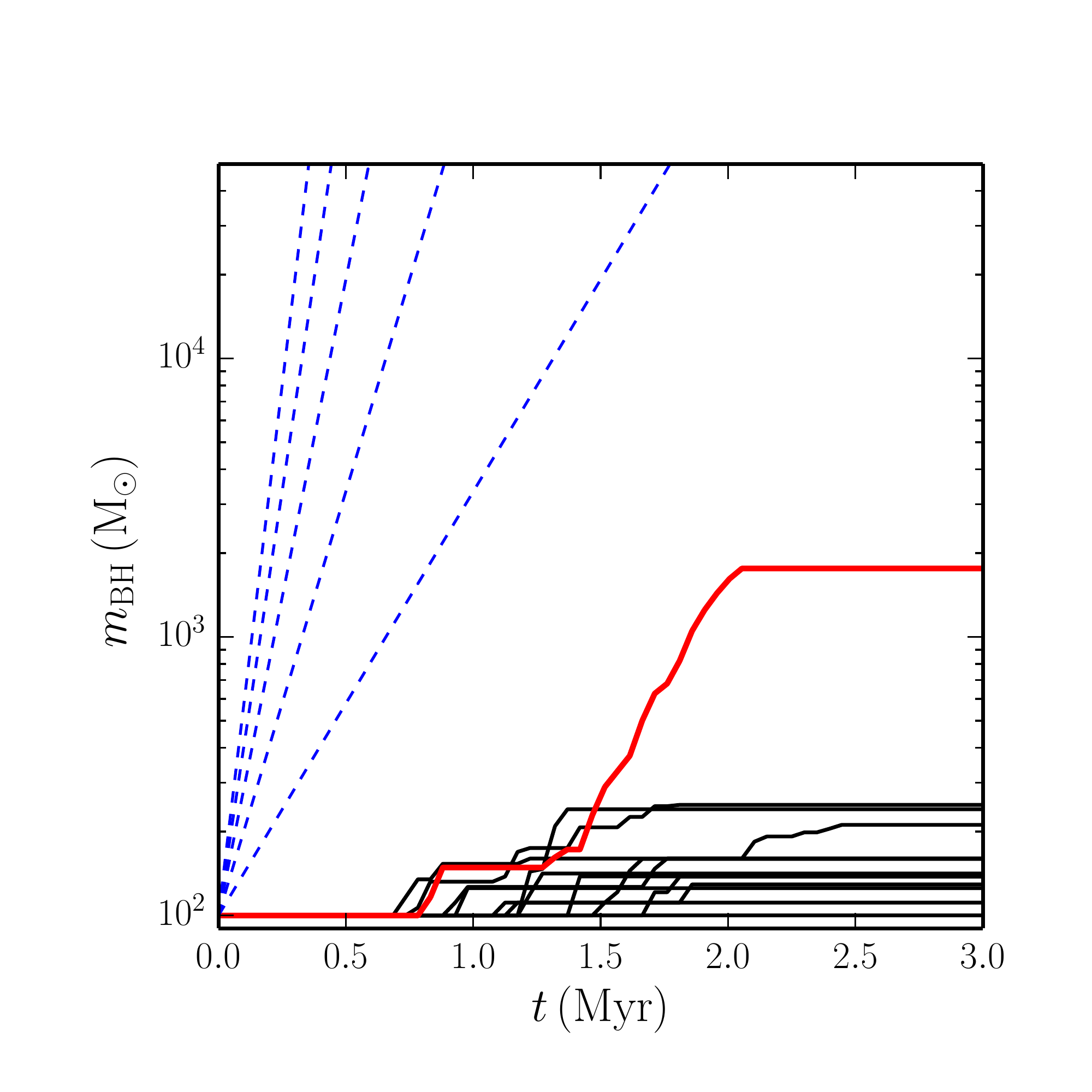}
  \includegraphics[width=0.3\textwidth]{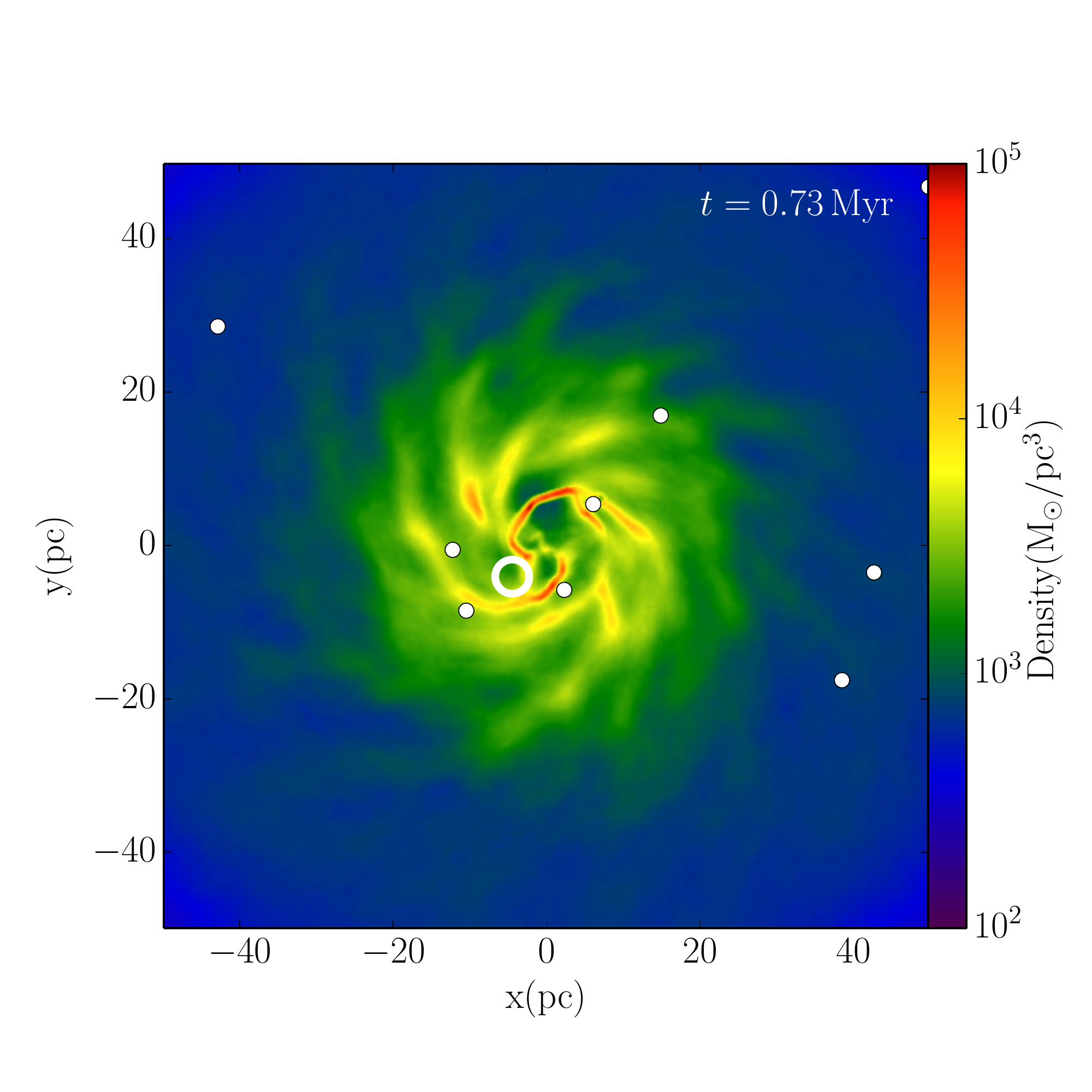}
  \includegraphics[width=0.3\textwidth]{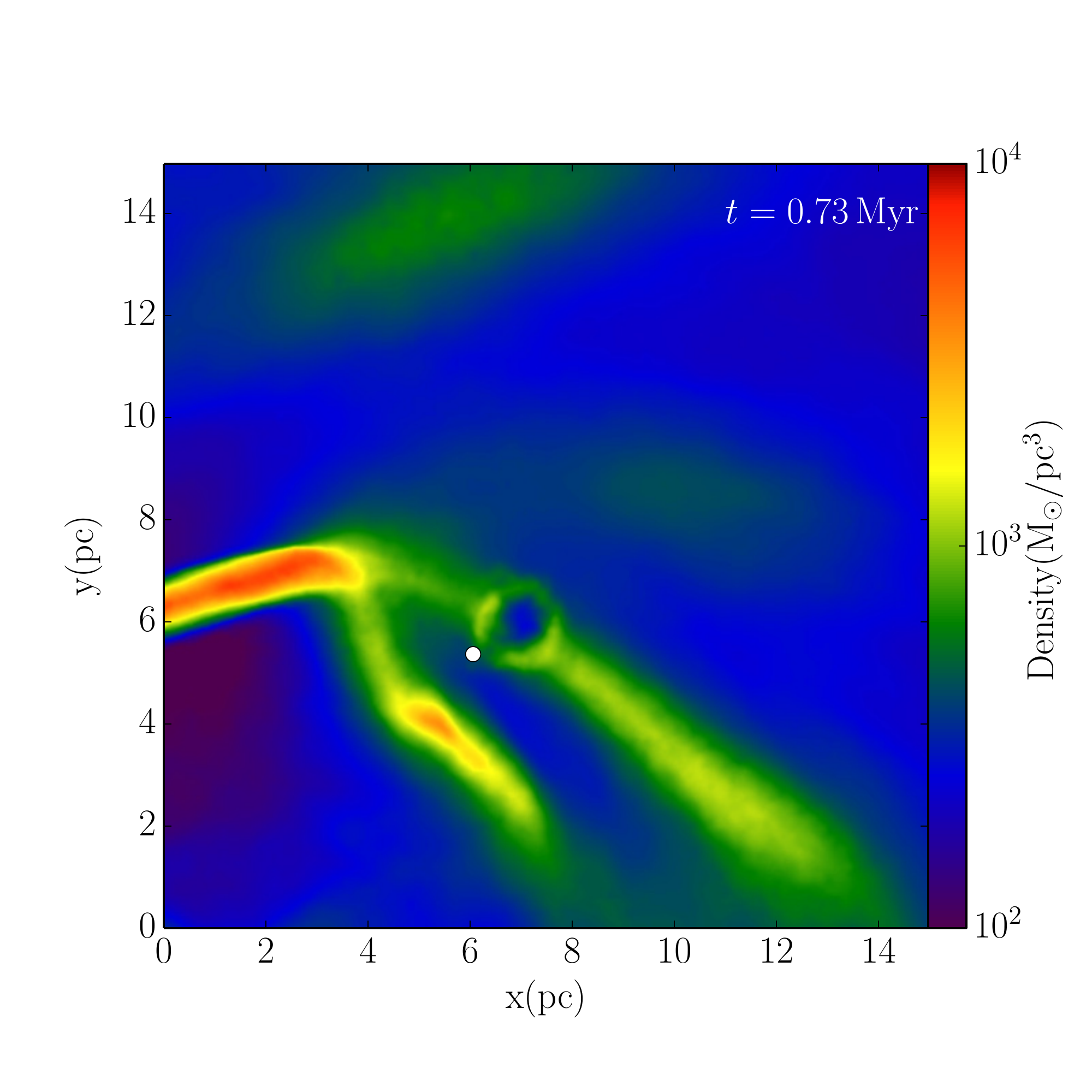}
\caption{\rm{Left panels: BH masses as a function of time for runs
    high\_G\_0.1 (top panel) and high\_G (bottom panel). The red
    lines correspond to the most massive BHs (BH$_{\rm top}$) at the
    end of the runs, while the blue dashed lines trace accretion
    histories at fixed Eddington ratios of 500, 400, 300, 200 and 100,
    respectively. Central panels: gas density maps for the
    two runs at $t=0.73$ Myr. Right panels: zoom in of a region heated by BH feedback. The white dots mark the
    positions of the BHs.}}
\label{fig:mtime_high}
\end{figure*}

In order to asses how numerical resolution affects our results, we analyse the two high resolution \gizmo runs (high\_G and high\_G\_0.1), and compare the outputs to the low resolution cases discussed above. Figure~\ref{fig:mtime_high} shows the accretion history of BHs (left panels) and the effect the accretion feedback has on the gas (right panels). Because of  the higher resolution we can now resolve a smaller accretion region around each BH, which has the net effect of reducing the BH mass growth compared to the corresponding low resolution runs. Nevertheless, it is apparent how, also in these high resolution simulations, BH mass growth is strongly suppressed in the radiatively efficient case (top panels). Indeed, for $\epsilon=0.1$, BH$_{\rm top}$ increases its mass by only $\simeq 50\%$ of its initial value. 
We stress again that the different radiative efficiency is only marginally responsible of the different accreted mass in the two cases. As clearly shown in the right panels of Figure~\ref{fig:mtime_high}, the
largest effect is played by the accretion feedback that, in the standard high-efficiency case, evacuates the region
closer to the BHs, hence inhibiting further gas accretion.

The implementation of a physically motivated radiative inefficient
accretion mode is then a necessary condition for a fast, highly
super--Eddington growth of BHs in our simulations, but, as we will show next, is not sufficient. 
In the following we will focus only on runs including the slim disc prescription, in
order to link episodes of super--Eddington growth with the physical
state of the BHs and of the nuclear disc, with the ultimate aim of
understanding the processes that can possibly lead to high accretion rates.

\begin{figure*}
\vspace{-0.5cm}
\hspace{-0.6cm}
\includegraphics[width=0.45\textwidth]{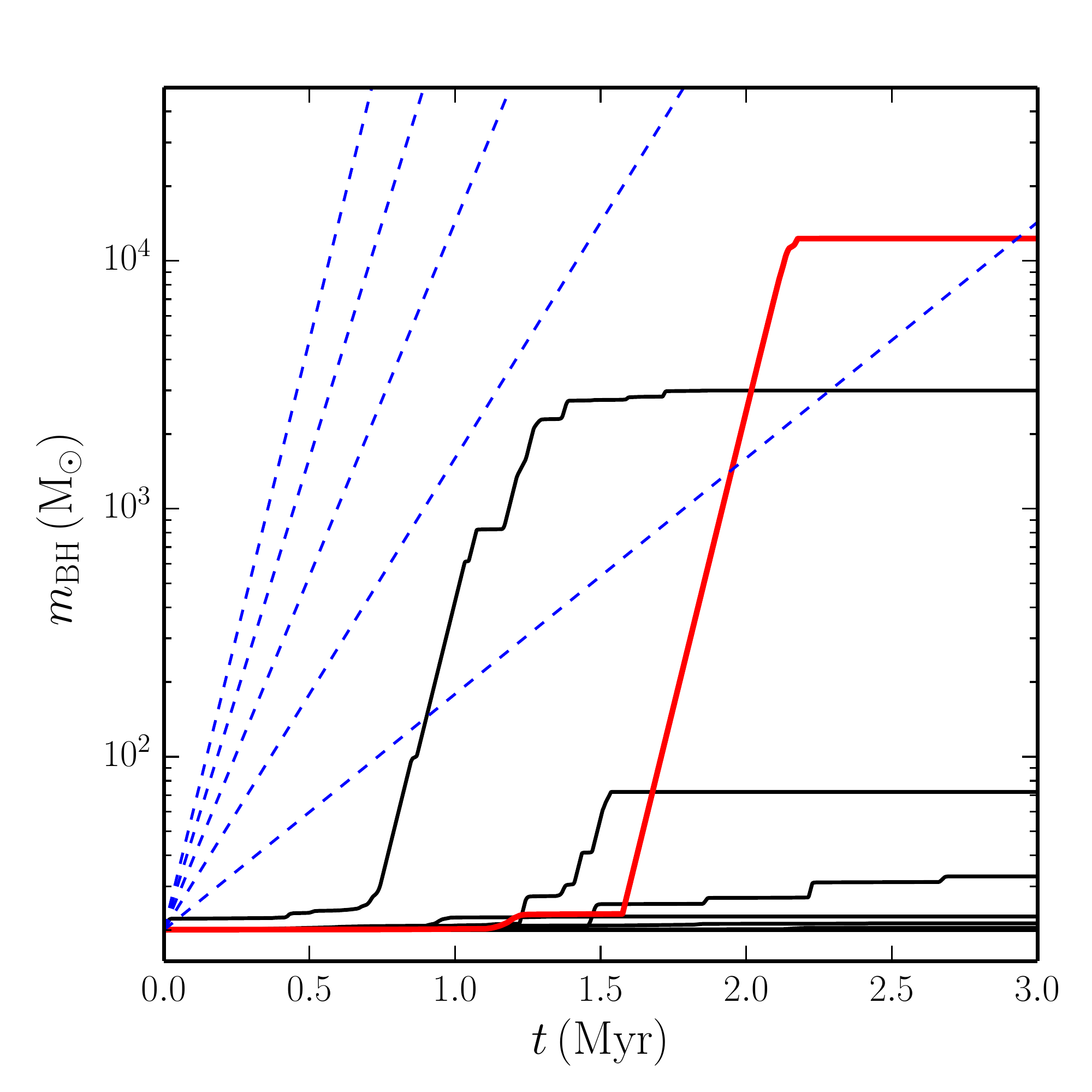}\hspace{0.44cm}
\includegraphics[width=0.45\textwidth]{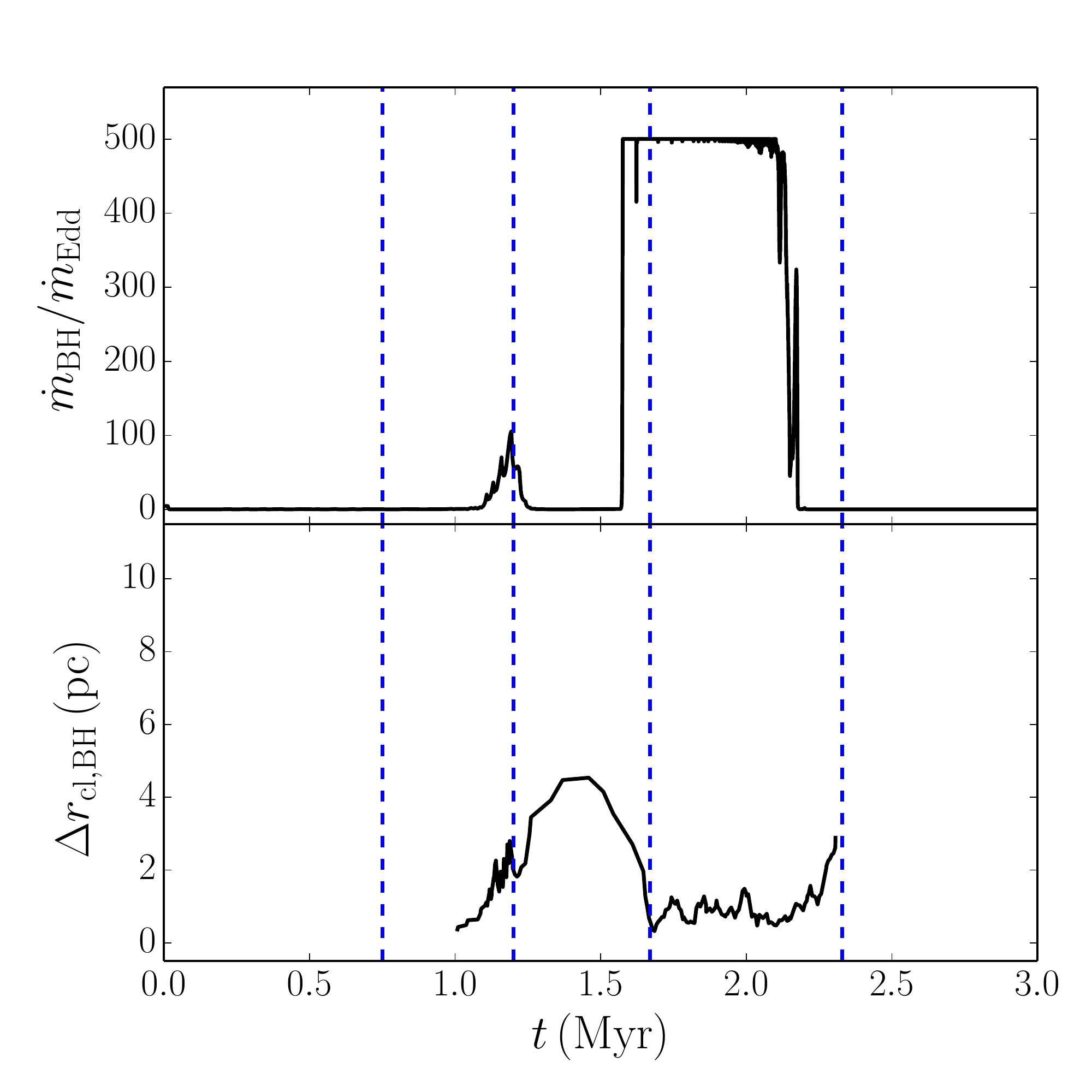}\\
\includegraphics[width=0.47\textwidth]{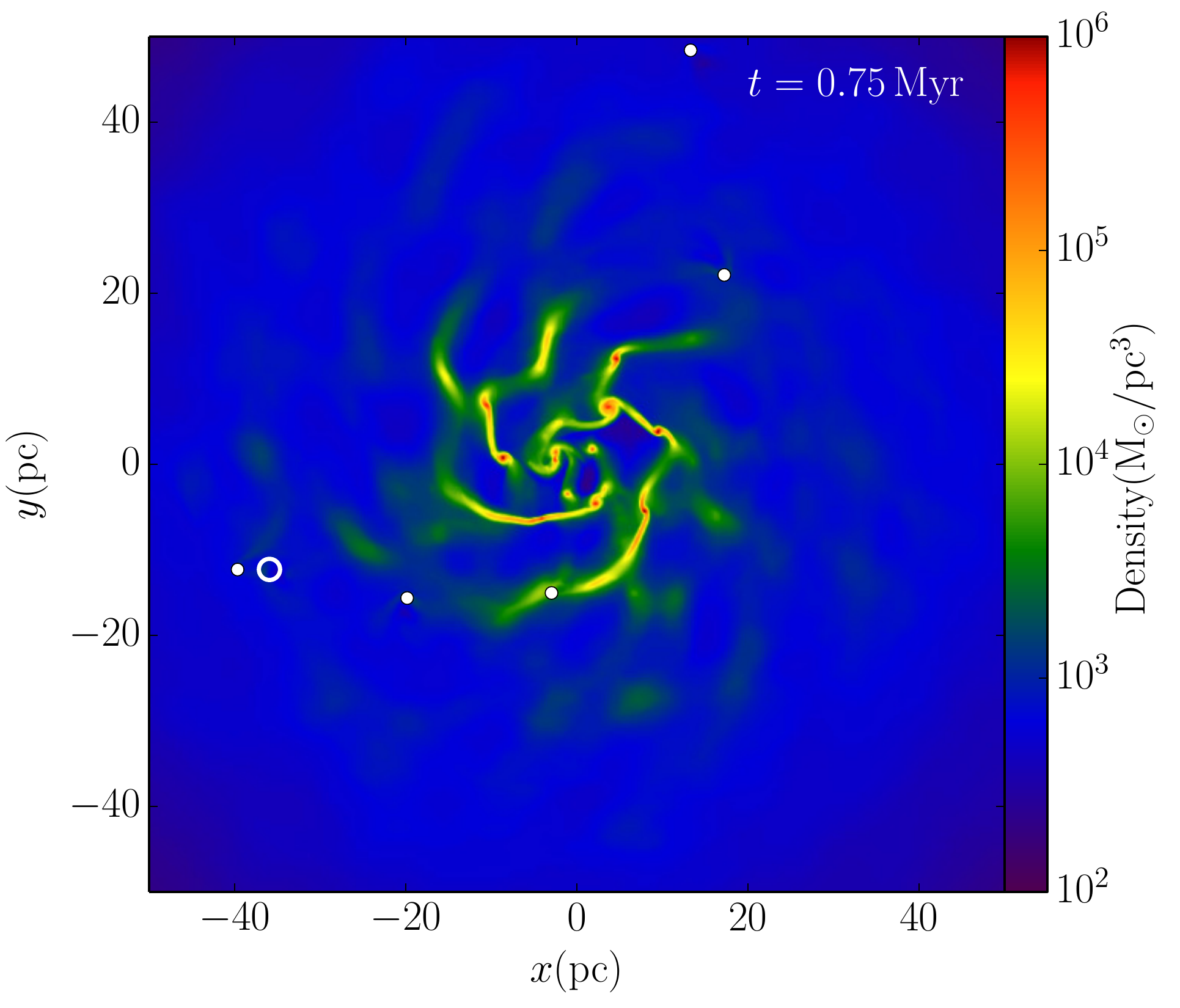}
\includegraphics[width=0.47\textwidth]{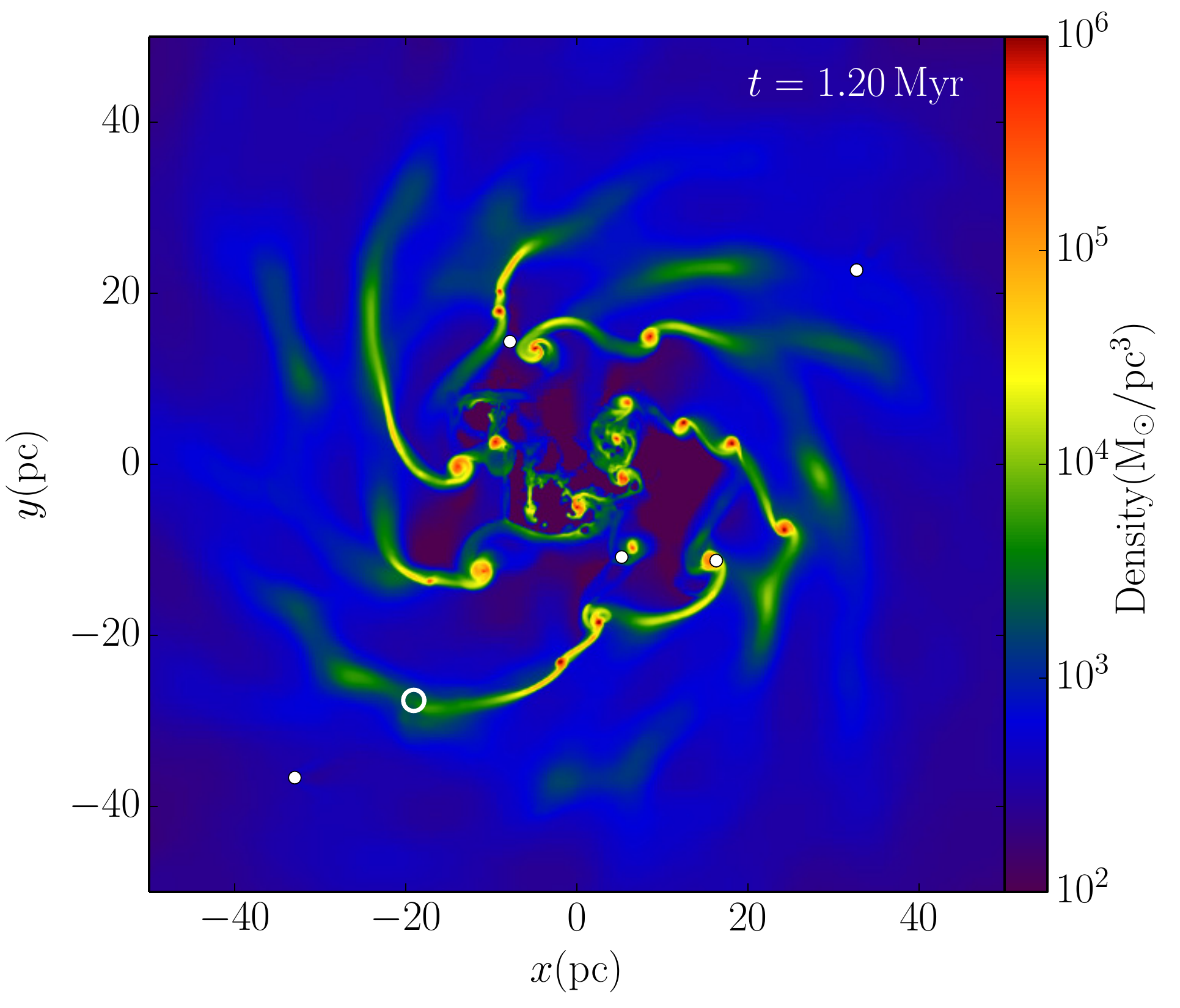}
\includegraphics[width=0.47\textwidth]{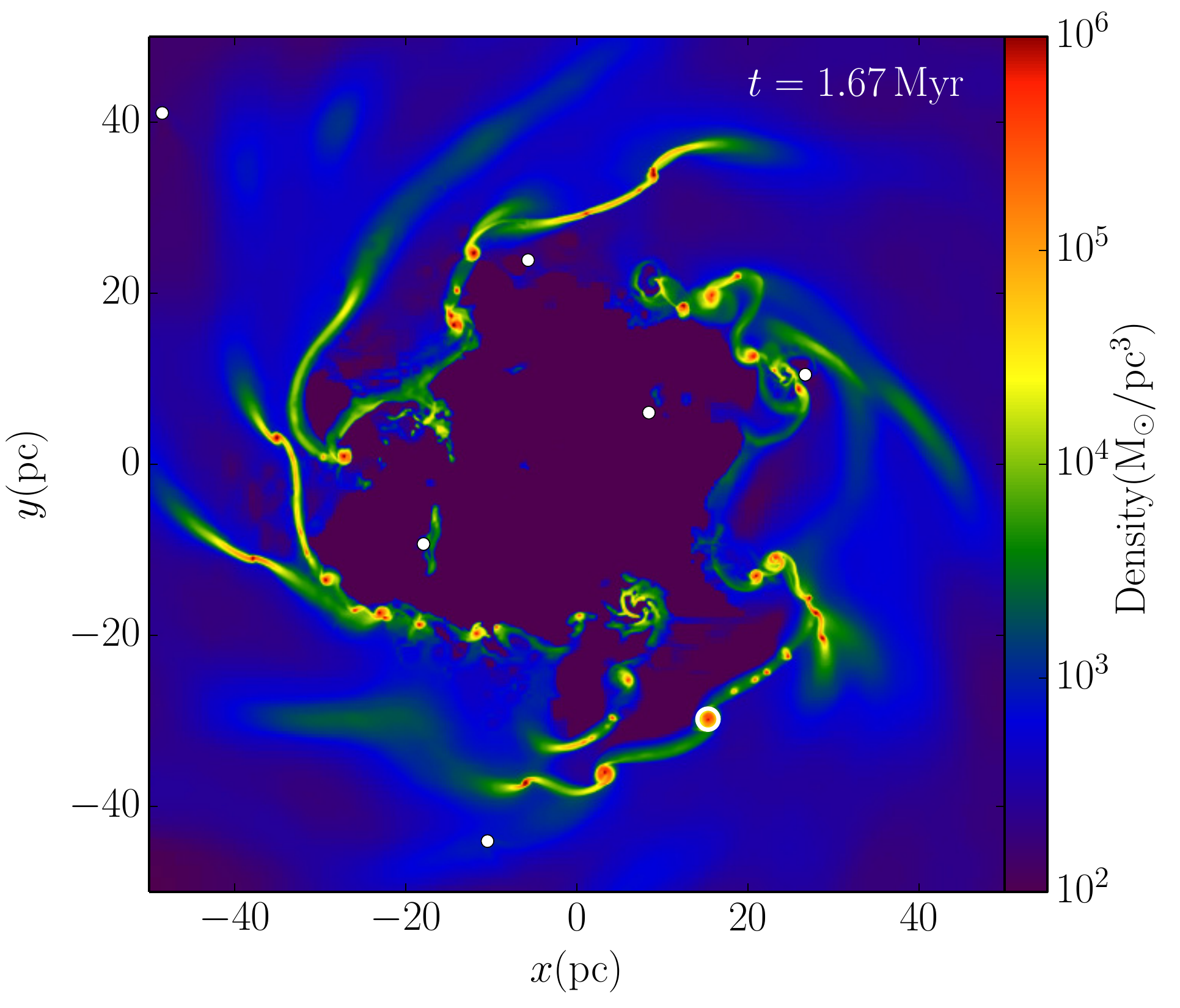}
\includegraphics[width=0.47\textwidth]{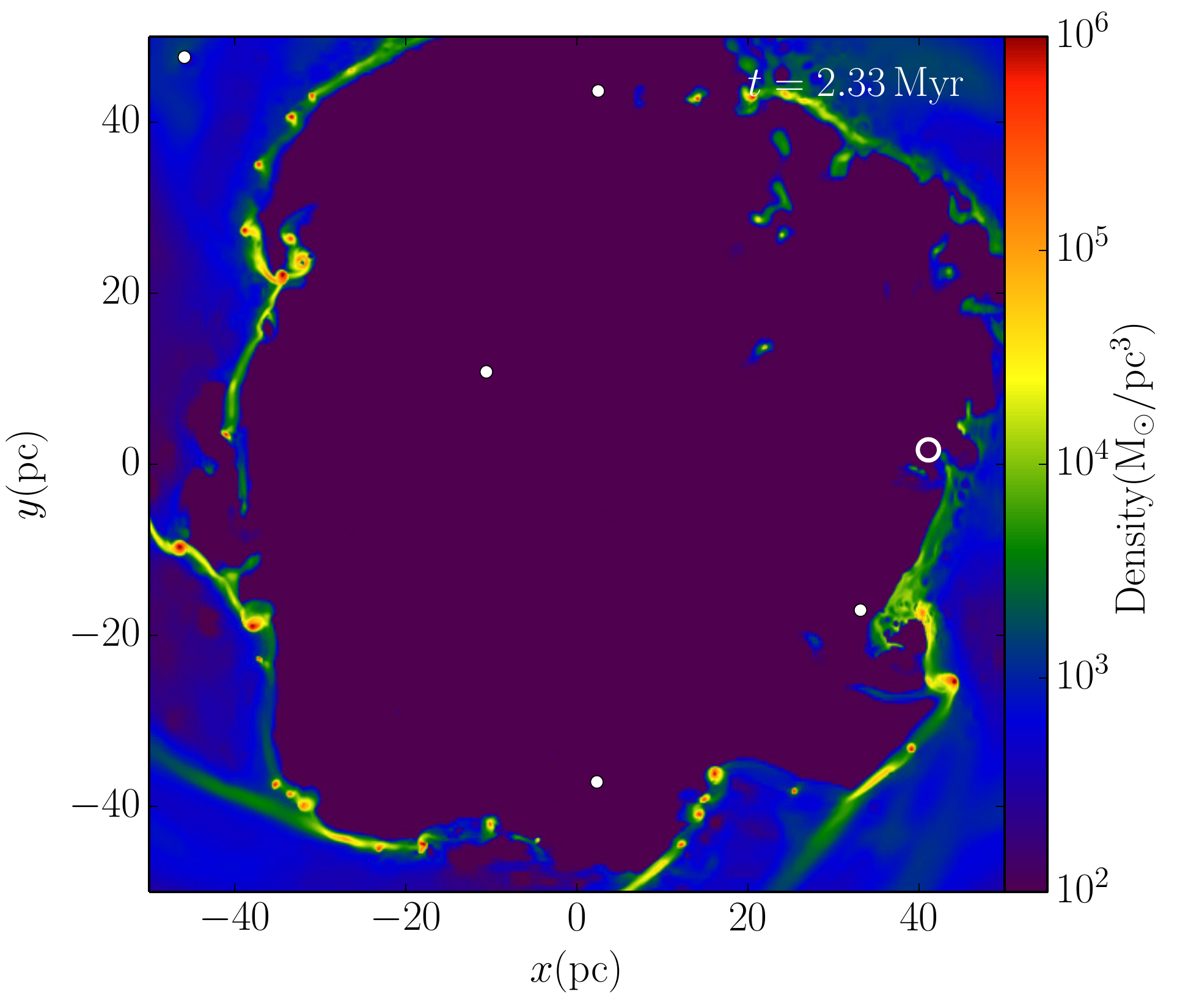}
\caption{\rm{Run med\_R. Upper left panel: BH masses vs time for
    all the 20 BHs. The dashed lines show the slope of accretion
    episodes at 500, 400, 300, 200 and 100 $\dot{M}_{\rm E}$. Upper
    right panel: accretion rate for BH$_{\rm top}$, and distance from
    the clump BH$_{\rm top}$ bounds to during the peak of its mass
    growth. Middle and lower panels show the density in the equatorial
    disc plane of the gas at $t=0.75$, 1.2, 1.67 and 2.33 Myr
    (corresponding to the times highlighted by the dotted lines in the
    upper upper right panel). The BH$_{\rm top}$ is reported as large white
    ring, while the other BHs are shown as smaller white dots.}}
\label{fig:mtime_Ramses_high}
\end{figure*}

Figure~\ref{fig:mtime_Ramses_high} shows the results of the highest
resolution RAMSES run med\_R. The upper left panel reports the
mass evolution of the 20 BHs as a function of time. As for the low\_G 
simulation discussed above, the implementation of the slim disc
efficiency prescription allows BH$_{\rm top}$ (shown as a red line) to grow within 3 Myr 
by up to $\sim 3$ orders of magnitude in mass. Note that BH$_{\rm top}$ is not necessary 
the earliest growing BH of the cluster. 

%We postpone the
%discussion about the resolution dependent differences in the BH
%accretion histories to the end of this section. 

The upper right panel
of figure~\ref{fig:mtime_Ramses_high} focuses on 
BH$_{\rm top}$ alone, showing the time evolution of the accretion rate, and the corresponding distance from
the gas clump the BH bounds to during the peak of its mass growth.
The accreting clump forms out of a spiral stream developing in the cooling disc,
and can not be clearly identified as a bound structure before $t\approx 1$ Myr, as 
shown in the middle left panel. BH$_{\rm top}$ passes a first time through
the overdense stream (middle right panel), and experiences a short
$\lsim 0.1$ Myr super--Eddington accretion episode, but the radial
component of its velocity quickly is large enough to displace it from the
overdensity (as observed in the $\dot m_{\rm BH}$ plot, upper right panel). 
As the clump grows in mass (up to a maximum of $\sim 3\times 10^4\msun$ in gas), the BH$_{\rm top}$  
feels its gravitational attraction, and is eventually captured by the clump. At this
time BH$_{\rm top}$ undergoes a longer ($\sim 0.5$ Myr) intense
super--Eddington accretion phase. Being the initially small BH
surrounded by an overwhelmingly large and cold gas cloud, the BH accretes
at the maximum rate allowed by the code (i.e. $500 \times
\dot{M}_{\rm E}$) until almost all gas is turned into stars. 
At this point BH$_{\rm top}$ (already grown by 3 order of magnitudes in mass), together with stars exploding as SNae, can evacuate
the residual gas condensation (lower right panel). 
Note that BHs (including BH$_{\rm top}$) accrete most of their mass  
from, essentially, a single dense clump they randomly come across during the dynamical evolution of the system.

\begin{figure*}
\includegraphics[width=0.33\textwidth]{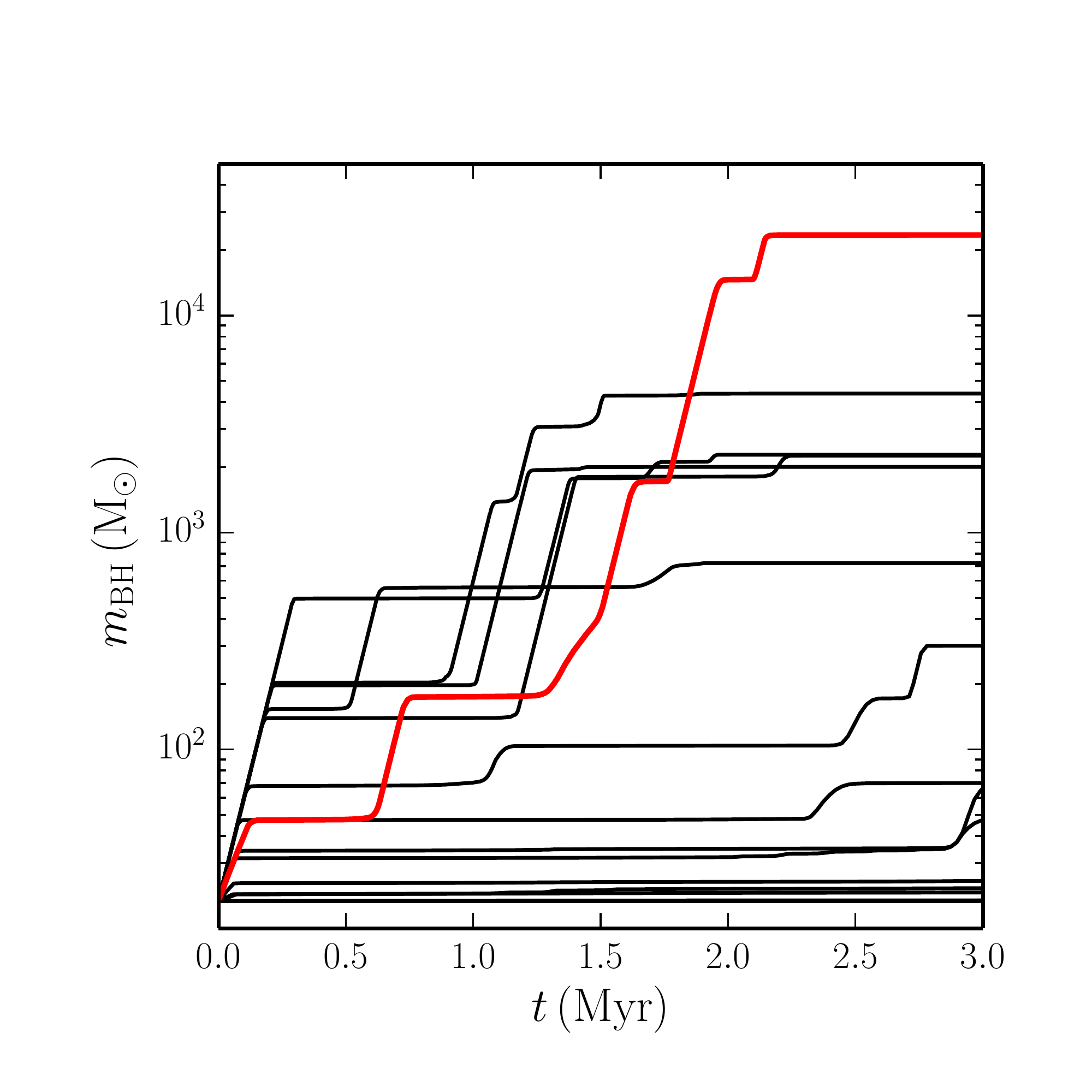}
\includegraphics[width=0.33\textwidth]{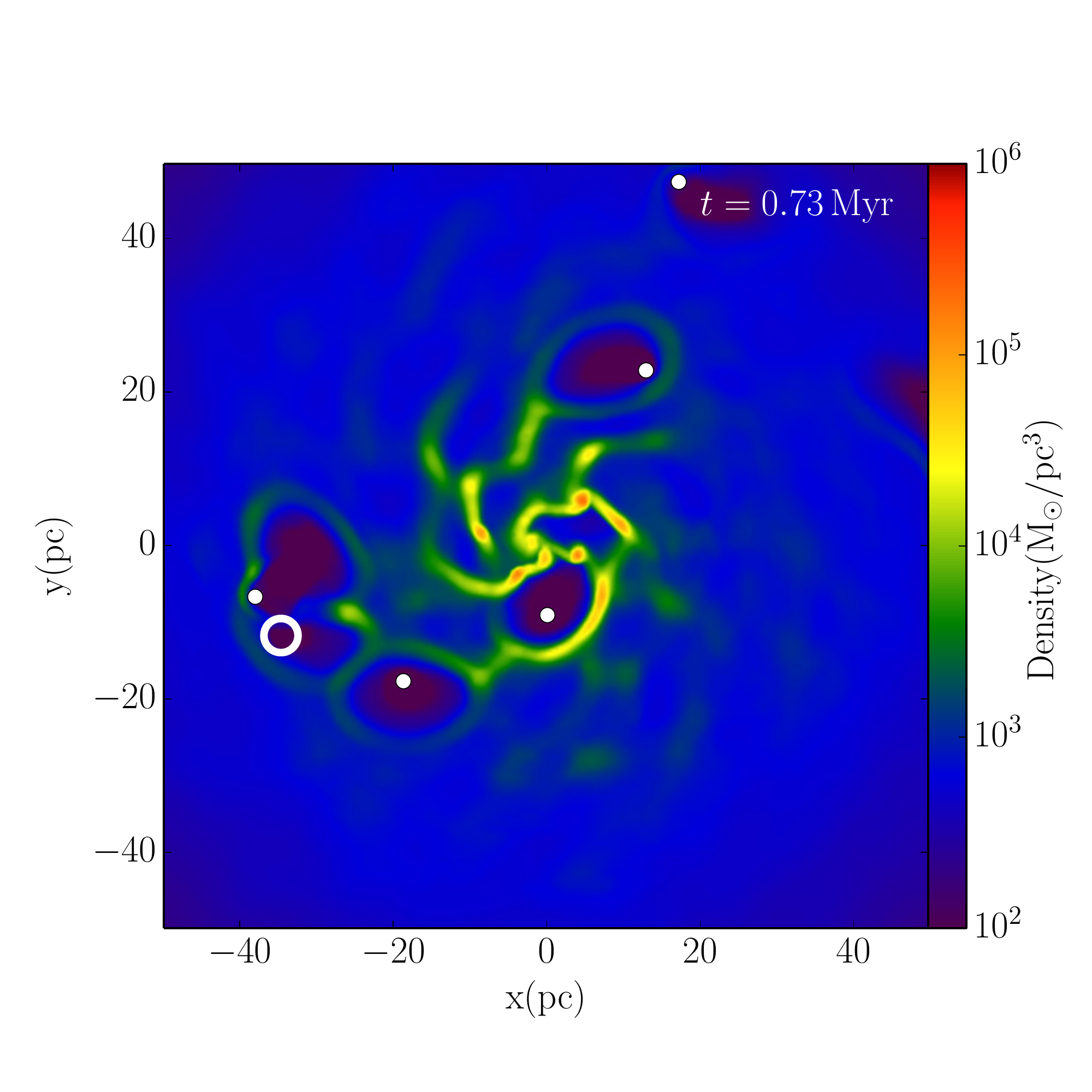}
\includegraphics[width=0.33\textwidth]{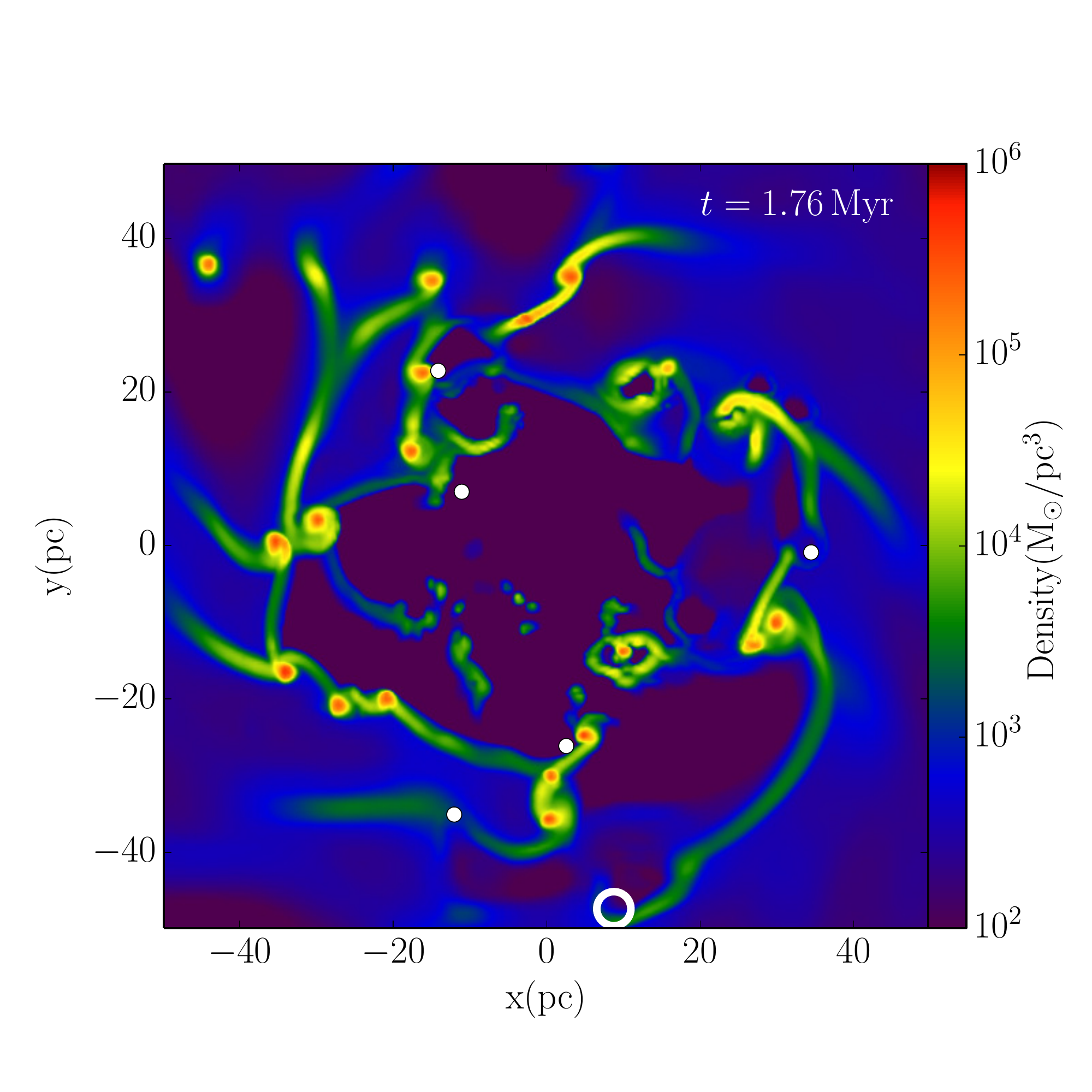}\\
\includegraphics[width=0.33\textwidth]{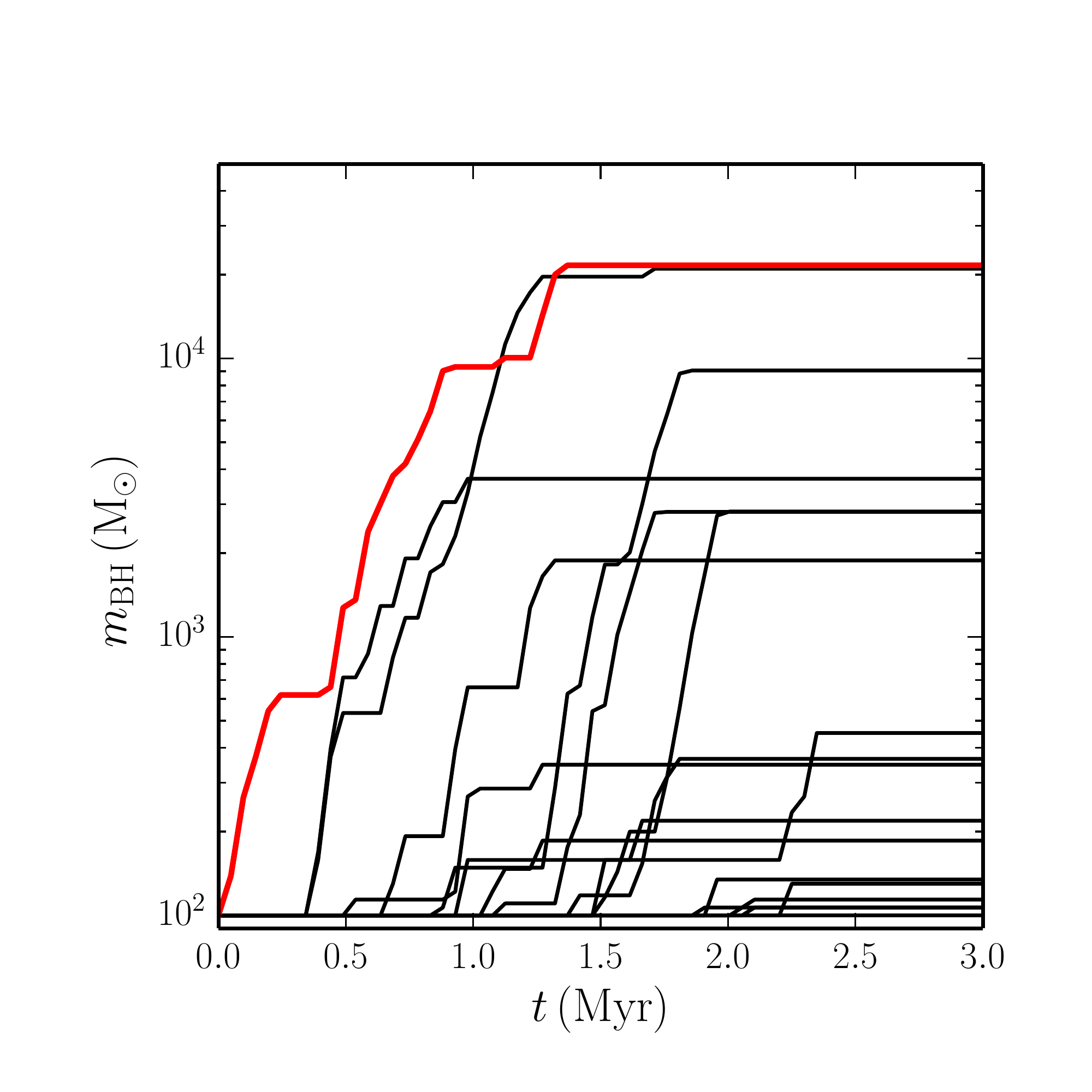}
\includegraphics[width=0.33\textwidth]{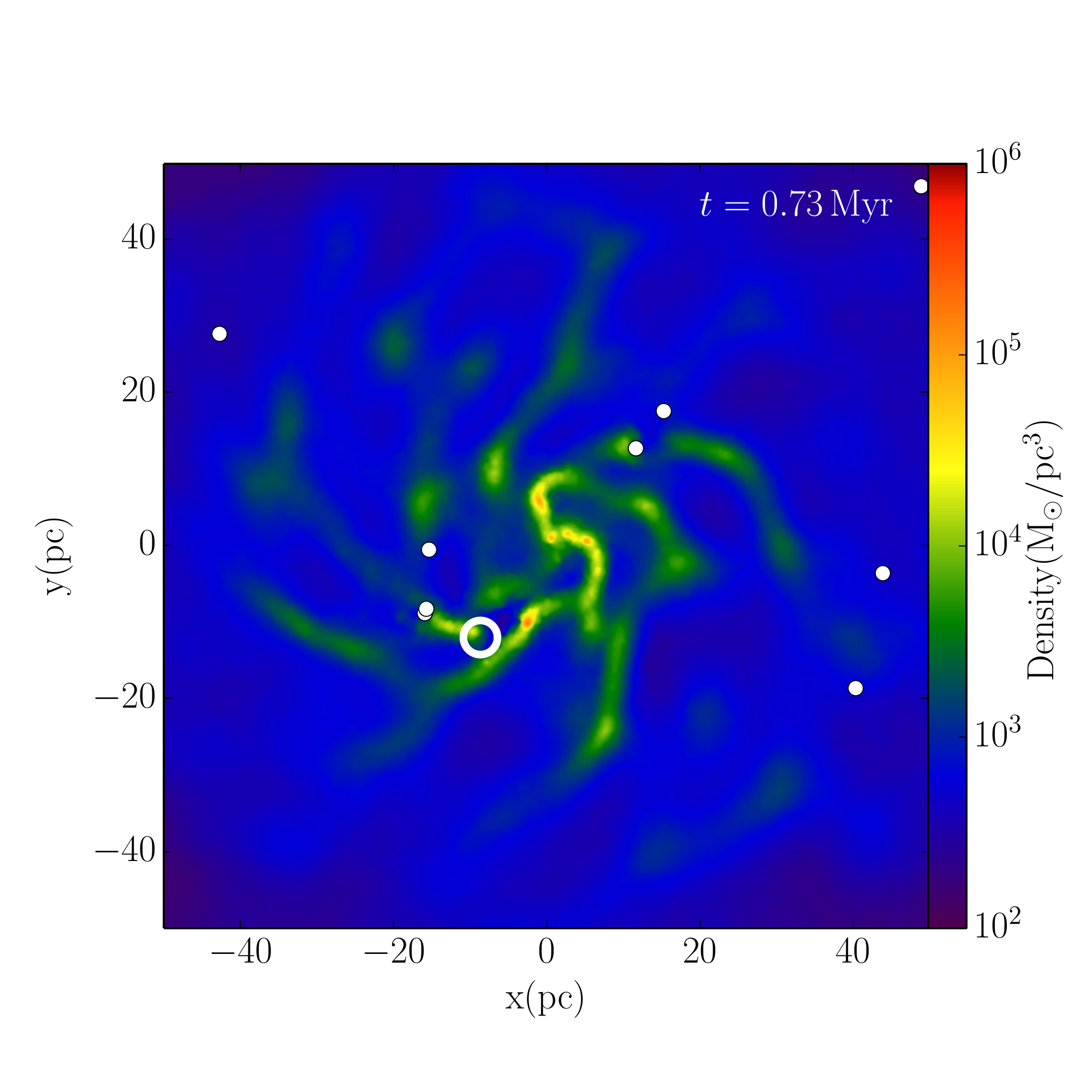}
\includegraphics[width=0.33\textwidth]{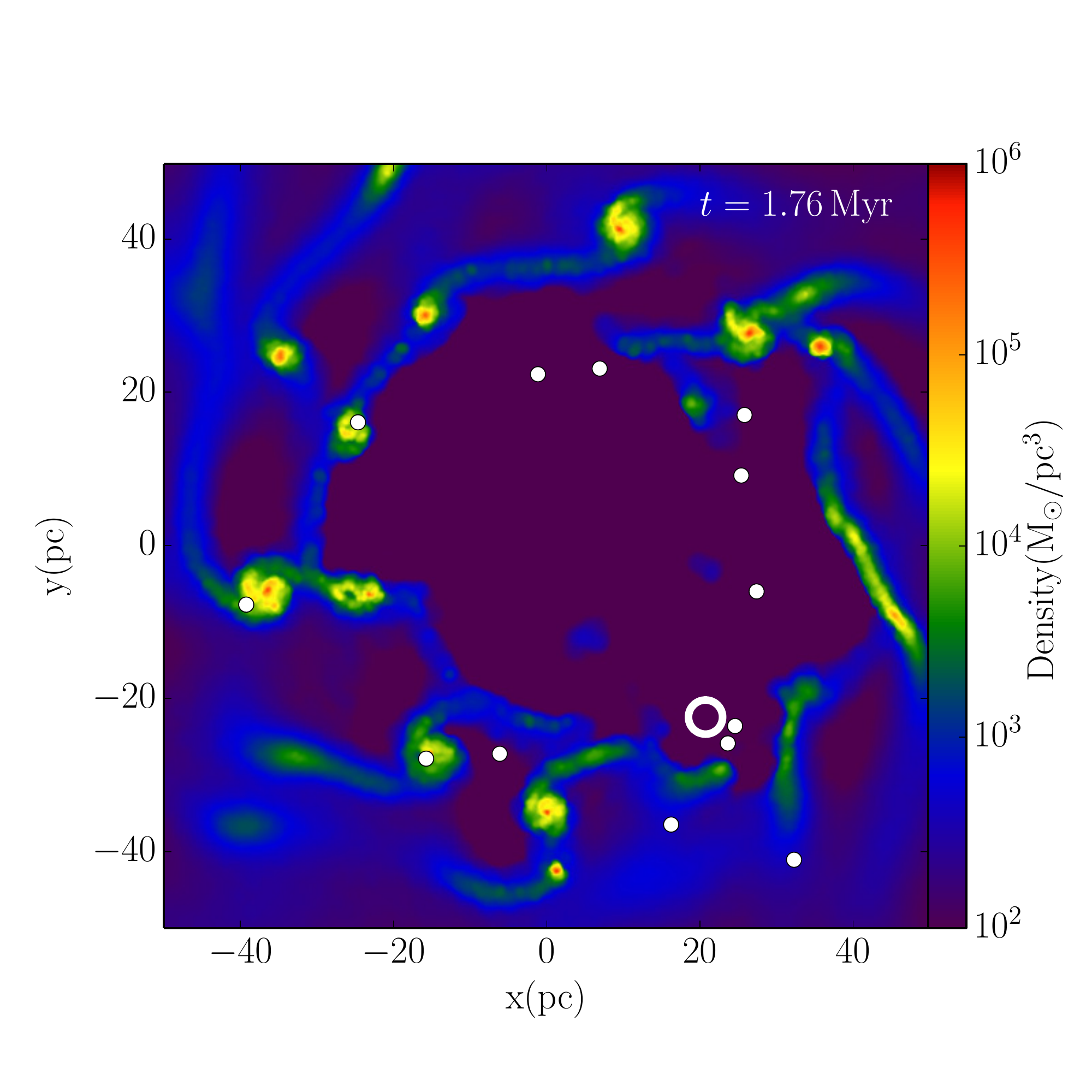}\\
\includegraphics[width=0.33\textwidth]{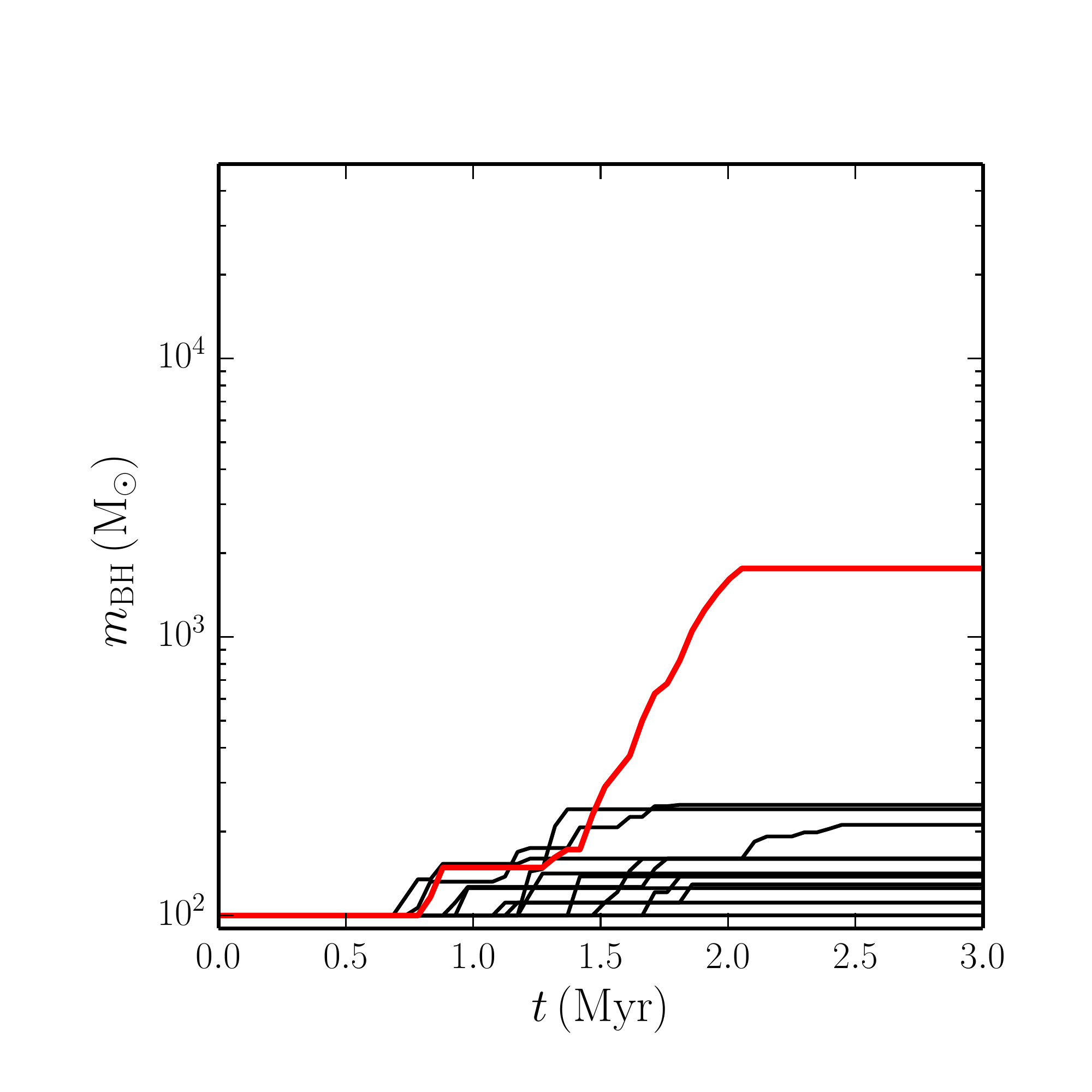}
\includegraphics[width=0.33\textwidth]{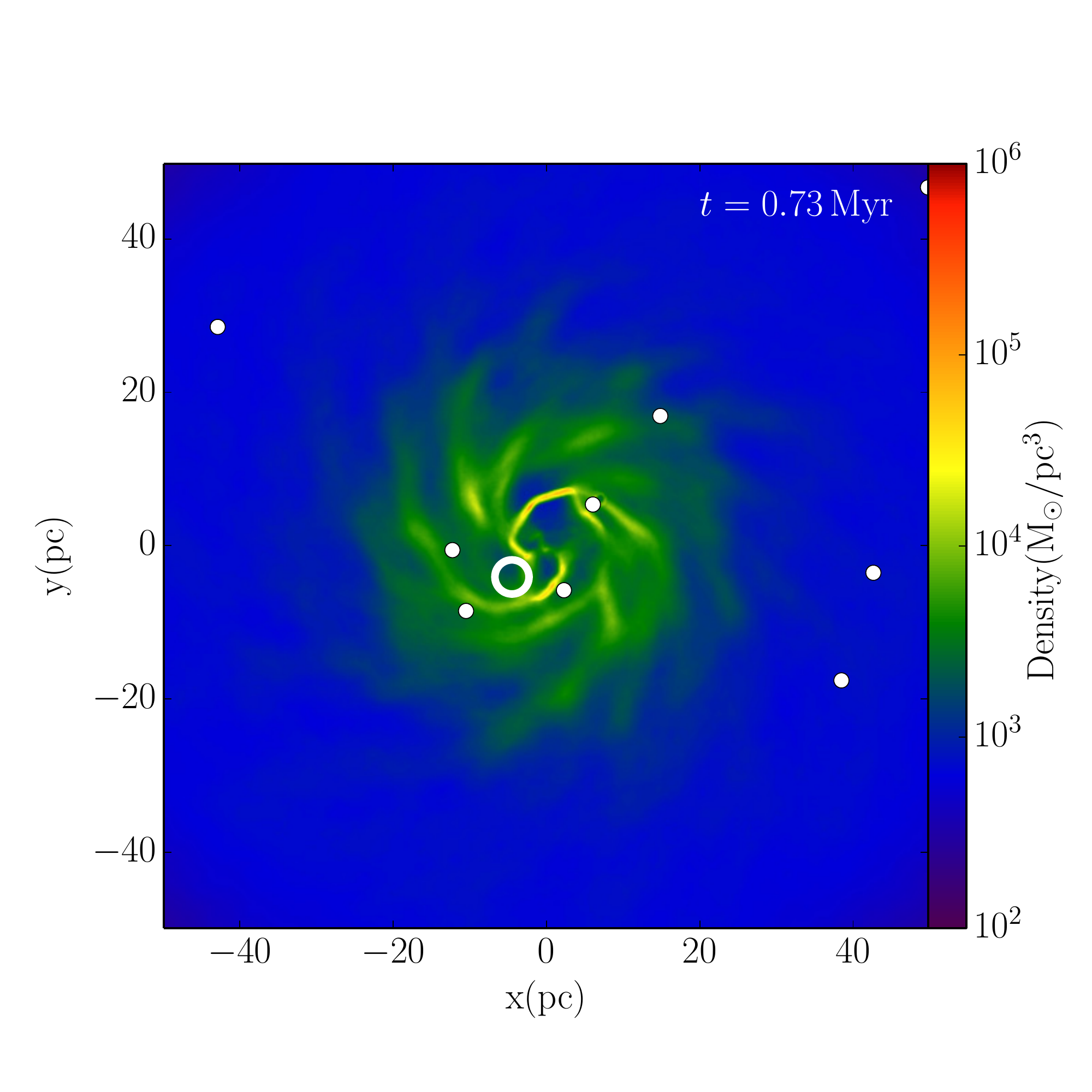}
\includegraphics[width=0.33\textwidth]{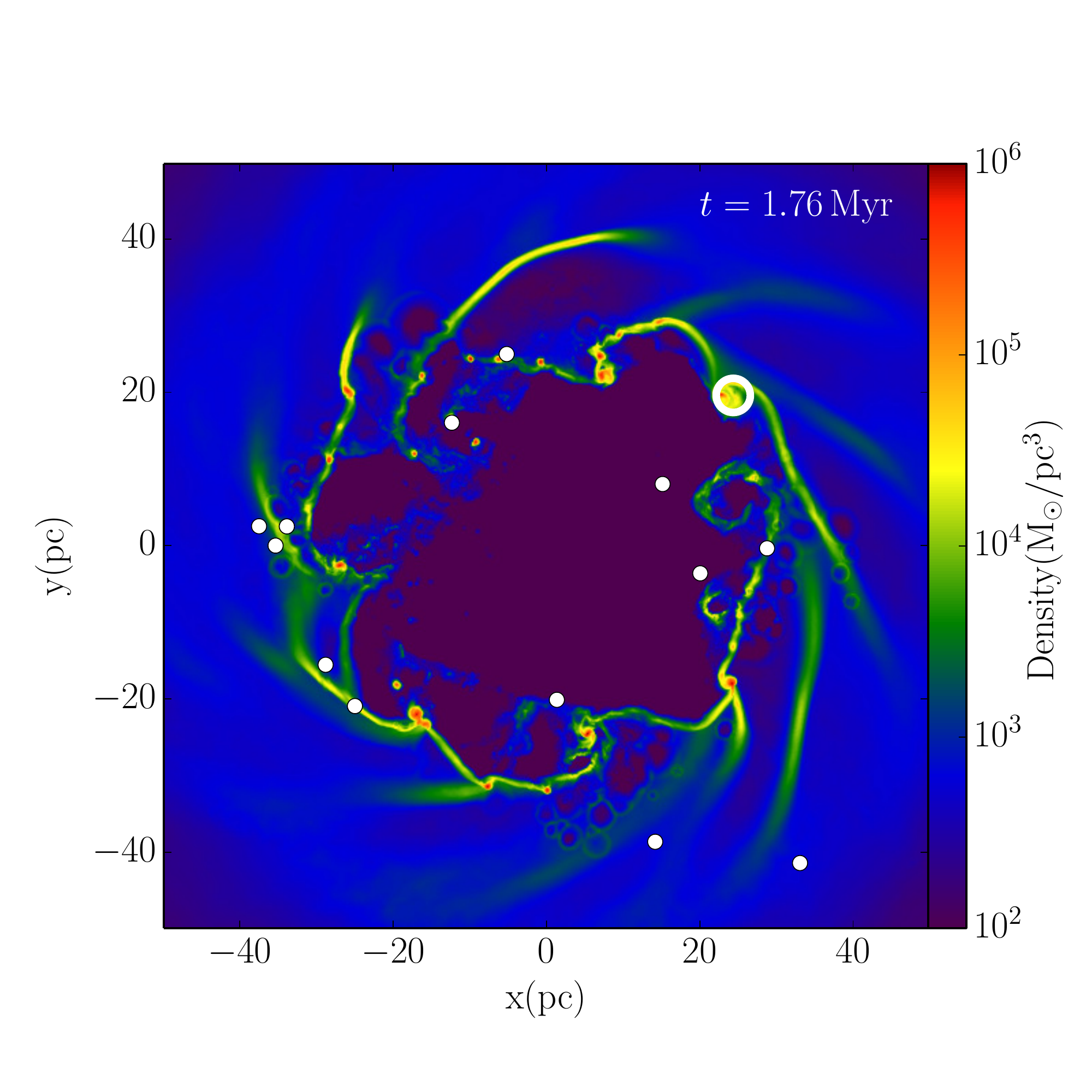}
\caption{\rm{Upper panels: Mass as a function of time of the BHs
    (left panel), gas density at $t= 0.73$ Myr (central panel) and
    $t=1.76$ Myr for the low\_R run.  The positions of the BHs are
    shown as white dots. The growth of BH$_{\rm top}$ is highlighted
    in the left panel with a red line, and its position in the central
    and right panels is marked with a large white ring.  Middle and
    lower panels, the same as the upper panels for run low\_G and
    high\_G, respectively.}}
\label{fig:map_med_hom}
\end{figure*}

It is important to realise that the gravitational capture of a BH by 
a dense gas clump is intrinsically stochastic, as clumps form in the disc via gravitational instabilities of
cooling gas independently of the presence of seed holes. 
While the BH--capture process is common in all the simulations we ran, 
the number and mass distributions of gas clumps and, consequently, the fraction of BHs that bind to them, 
in fact depend upon the spatial and mass resolution we achieve. 
Figure~\ref{fig:map_med_hom} shows a
comparison between runs with different spatial resolution. Among the
runs including the slim disc implementation, only run med\_R
(already shown in Figure~\ref{fig:mtime_Ramses_high}) is left out of
the direct comparison.

A first clear difference is observable at early times. The runs with
lower resolution show a faster initial growth of each individual
BH, and the number of growing BHs right after the beginning of the
runs ($t\lsim 0.5$ Myr) also increases with decreasing resolution.
These trends are caused by the larger accretion radius implemented in
the lower resolution runs. In these simulations the BHs can start
accreting well before the disc develops any significant
overdensity. For this reason the feedback of the early BH accretion
onto the gas is more efficient, as a larger energy is injected
in a lower density medium. As the resolution increases and the
accretion radius can be decreased, fewer BHs have an early start, as
in the med\_R run (upper left panel of
Figure~\ref{fig:mtime_Ramses_high}) and, more evidently, in the
high\_G run (lower left panel of Figure~\ref{fig:map_med_hom}). 

\begin{figure}
\includegraphics[width=0.45\textwidth]{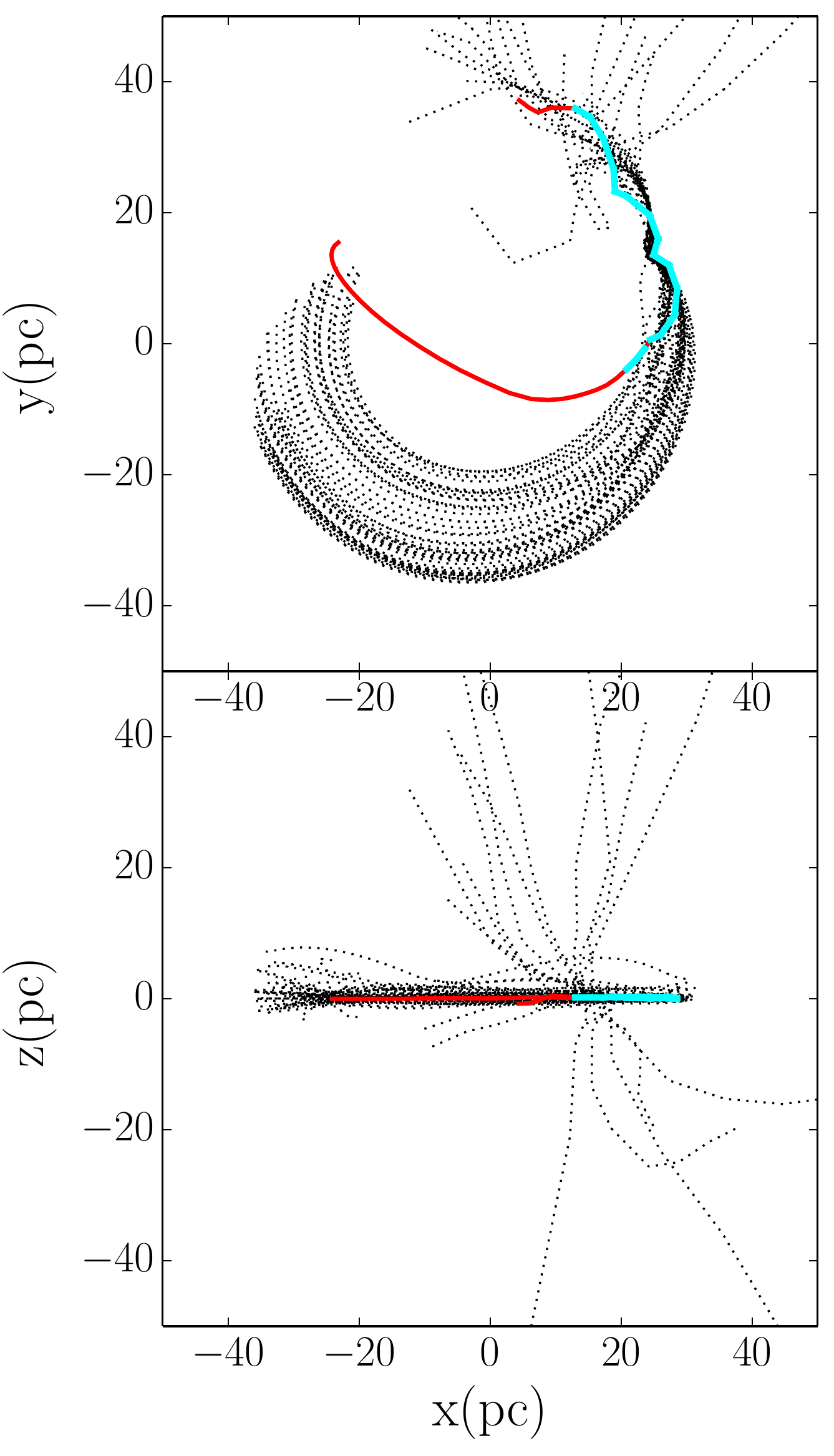}
\caption{\rm{Solid red lines show the face--on (top panel) and
    edge--on (bottom panel) projections of the trajectory of BH$_{\rm
      top}$ in run high\_G. Black dotted lines trace the orbits of a
    sample of the gas particles forming the gas clump BH$_{\rm top}$
    binds to. The accretion burst due to the  BH$_{\rm top}$--clump interaction is highlighted
in cyan.}}
\label{fig:clump}
\end{figure}

The high\_G run, thanks to the exquisite mass and spatial resolutions
achieved, shows a richness of structures observable directly in the
density map (see the lower right panel of Figure~\ref{fig:map_med_hom}
in particular), in which the formation of dense clumps as well as the
feedback exerted by the ongoing SF are clearly
visible. The gas particles tracing the gas evolution allow us to
follow the formation of the massive clump from which BH$_{\rm top}$
gains its mass. Figure~\ref{fig:clump} reports two different projections of
BH$_{\rm top}$ orbit along with the trajectories of 50 gas
particles randomly extracted from those forming the massive clump 
BH$_{\rm top}$ binds to and accretes from. The clump formation clearly
proceeds out of a gas gravitational instability within the dense disc,
and starts interacting with BH$_{\rm top}$ only when their orbits
intersect. Strong gravitational perturbations to the BH orbit are
clearly seen when the two systems bind gravitationally. The BH growth
then exerts a feedback onto the gas particles, that, together with stars exploding as SNae, results in a
%as a consequence,
%are partially ejected 
partial ejection of particles from the BH neighbourhoods and out of the disc
plane (as clearly see in Figure~\ref{fig:clump} lower panel). 

\begin{figure*}
\centering
\includegraphics[scale=0.5]{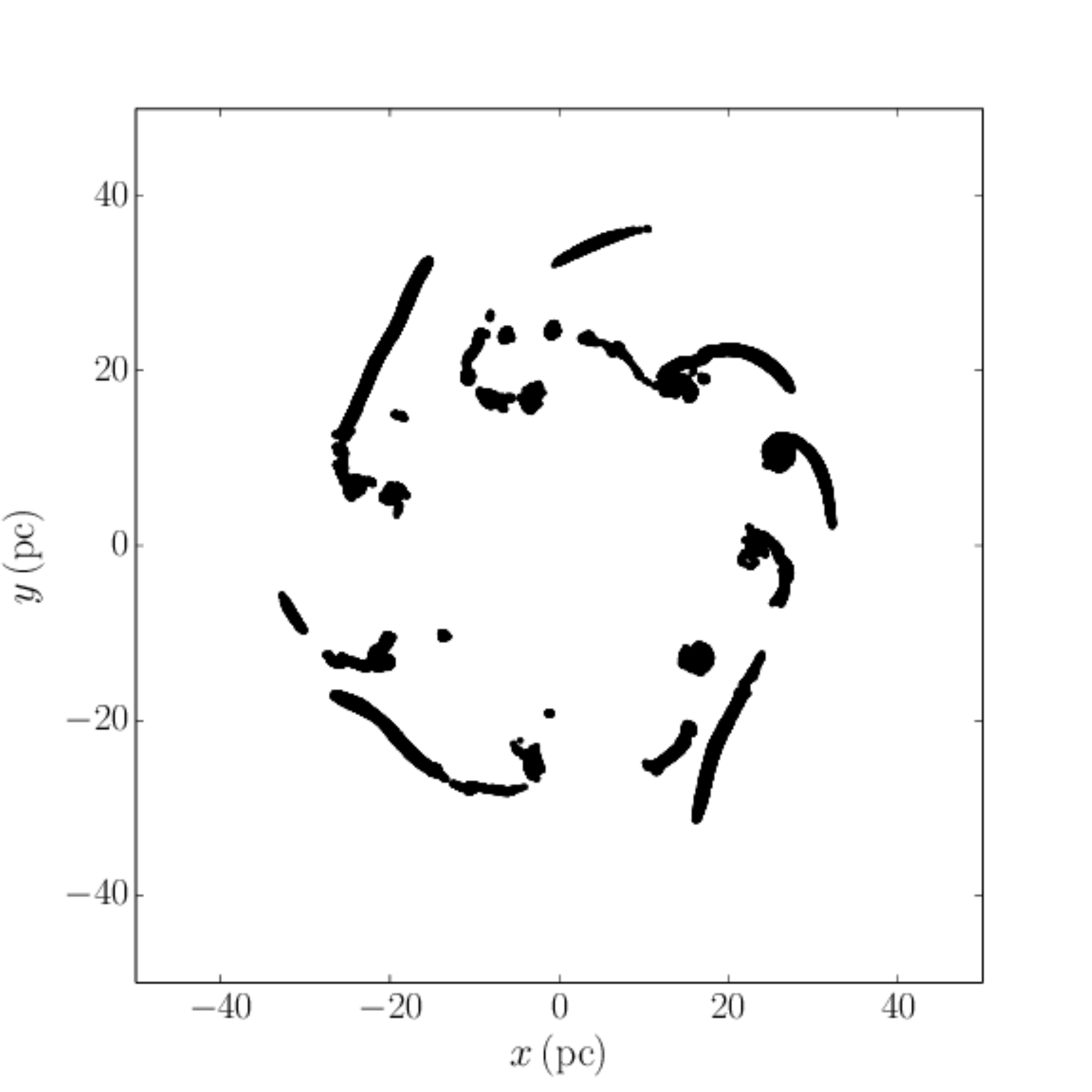}
\caption{Map of clumps in the high\_G run at $t \sim 1.6$ Myr. The clumps can be grouped in two different classes: CL exhibit a disc-like shape and BS are extended bound streams.}
\label{fig:clumps}
\end{figure*}

We finally analyse the physical properties of clumps in the high\_G run using the public tool SKID\footnote{\url{https://github.com/N-BodyShop/skid}}, and report our results for both the disc-like clumps  and the more extended gravitationally bound streams in table~\ref{tab:clumps}. Clump properties are evaluated at $t\sim 1.6$ Myr,  when $BH_{\rm top}$ is in its main accretion phase. Fig.~\ref{fig:clumps} shows the clumps in the disc density map. The reported circular velocity is computed at the half mass radius of the clump. It is worth noticing that typical masses 
(ranging from few $10^2 \msun$ up to few $10^5\msun$), as well as 
the density of our clumps are compatible with the observed properties of local giant molecular clouds (GMCs), though the latter are much less compact \citep {lombardi10}. The average temperature of the clumps is much higher than the typical temperature of GMCs, mainly because we neglected molecular hydrogen cooling and, at the same time, the metal production in the few Myr of our simulations is not sufficient to significantly alter the gas cooling function. While allowing for a more efficient cooling would initially boost SF in the clumps, further SF could be hampered by the enhanced UV dissociating flux \footnote{We can speculate that, if clumps become self-shielded to dissociating radiation and SF continues unimpeded, dense stellar systems could form, which eventually might be prone to dynamical instabilities. A detailed analysis of the formation of dense stellar clusters in our simulations is beyond the scope of the current study.}.

\begin{table*}
\begin{tabular}{ccccc}
$M^a$ & $R_{1/2}^b$ & $v_{1/2}^c$ & $T^d$& $\rho^e$ \\
\hline
\hline
\multicolumn{5}{c}{Bound clumps} \\
\hline
$8.70\times 10^{2}$ & $ 0.21$ & $ 2.98$ & $ 8.77\times 10^{3} $ & $6.77\times 10^{5}$\\
$1.80\times 10^{3}$ & $ 0.31$ & $ 3.51  $ & $ 8.89\times 10^{3} $ & $7.06\times 10^{5}$\\
$2.53\times 10^{3}$ & $ 0.24$ & $   4.77$ & $ 8.65\times 10^{3} $ & $7.73\times 10^{5}$\\
$1.68\times 10^{4}$ & $ 0.16$ & $ 14.96$ & $ 9.94\times 10^{4} $ & $2.26\times 10^{7}$\\ $3.21\times 10^{4}$ & $ 0.28$ & $ 15.85$ & $ 2.91\times 10^{4} $ & $5.94\times 10^{6}$\\
$5.58\times 10^{4}$ & $ 2.48$ & $ 6.95  $ & $ 1.64\times 10^{4} $ & $2.71\times 10^{6}$\\
$6.75\times 10^{4}$ & $ 0.27$ & $ 23.22$ & $ 5.78\times 10^{4} $ & $1.31\times 10^{7}$\\
$8.05\times 10^{4}$ & $ 0.48$ & $ 18.98$ & $ 3.47\times 10^{4} $ & $7.04\times 10^{6}$\\
$8.87\times 10^{4}$ & $ 0.48$ & $ 20.04$ & $ 2.95\times 10^{4} $ & $5.73\times 10^{6}$\\
$9.38\times 10^{4}$ & $ 0.41$ & $ 22.27$ & $ 4.12\times 10^{4} $ & $8.40\times 10^{6}$\\
$1.24\times 10^{5}$ & $ 0.91$ & $ 17.07$ & $ 2.23\times 10^{4} $ & $4.09\times 10^{6}$\\
$1.60\times 10^{5}$ & $ 0.70$ & $ 22.14$ & $ 3.96\times 10^{4} $ & $8.18\times 10^{6}$\\
$2.14\times 10^{5}$ & $ 1.66$ & $ 16.64$ & $ 1.55\times 10^{4} $ & $2.51\times 10^{6}$\\
\hline
\multicolumn{5}{c}{Bound streams}\\
\hline
$1.13\times 10^{4}$ & $ 0.89 $ & $ 5.24  $ & $ 8.24\times 10^{3} $ & $5.45\times 10^{5}$\\
$3.51\times 10^{4}$ & $ 1.59$ & $   6.89$ & $ 8.43\times 10^{3} $ & $5.63\times 10^{5}$\\
$8.28\times 10^{4}$ & $ 2.55$ & $ 8.36  $ & $ 2.28\times 10^{4} $ & $2.48\times 10^{6}$\\
$1.70\times 10^{5}$ & $ 3.48$ & $ 10.24$ & $ 9.31\times 10^{3} $ & $ 8.05\times 10^{5}$\\
$1.73\times 10^{5}$ & $ 1.85$ & $ 14.16$ & $ 2.41\times 10^{4} $ & $ 4.60\times 10^{6}$\\
$2.15\times 10^{5}$ & $ 1.16$ & $ 20.02$ & $ 3.60\times 10^{4} $ & $7.54\times 10^{6}$\\
$3.25\times 10^{5}$ & $ 7.00$ & $ 9.99  $ & $ 1.92\times 10^{4} $ & $3.30\times 10^{6}$ \\
$3.90\times 10^{5}$ & $ 5.36$ & $ 12.51$ & $ 1.20\times 10^{4} $ & $1.53\times 10^{6}$\\
$4.83\times 10^{5}$ & $ 2.91$ & $ 18.91$ & $ 1.90\times 10^{4} $ & $3.29\times 10^{6}$\\
 
\hline
\hline
\end{tabular}
\caption{Properties of the clumps in the high\_G simulation at $t\sim 1.6$ Myr. The bound structures with an almost spherical shape have been classified as bound clumps, while the most asymmetric ones have been classified as bound streams. $~^a:$ clump mass in solar masses. $~^b:$ clump half mass radius in pc. $~^c:$ circular velocity at half mass radius in km/s. $~^d:$ gas mean temperatures in K. $~^e:$ gas mean number density in 
cm$^{-3}$.}
\label{tab:clumps}
\end{table*}

%Further and more detailed investigations are necessary to assess the most realistic conditions in high redshift gas rich galaxies.

We conclude by noticing that in our idealised runs the growth of the BHs
is finally halted by the star formation--driven gas consumption, and by
gas ejection triggered by SNae. 
However, in a cosmological perspective,
the galaxy nucleus would be replenished of gas coming from large
scale filaments and/or galaxy mergers. 
The very short duration of the
super--Eddington accretion bursts allows for the growth of stellar mass
BHs up to $\gsim 10^4 \msun$ or more on a time comparable (or even
shorter) than the star--formation timescale.
 
\section{Discussion and Conclusions}

We presented the results from a suite of numerical high resolution simulations aimed at studying
the accretion of stellar mass BHs in nuclear gaseous discs. We
implemented a new BH thermal feedback prescription, that takes into account
the possible occurrence of radiatively inefficient accretion bursts
during which the BHs can actually increase their masses at a
significantly super--Eddington pace. We have employed both AMR and
Lagrangian mesh-free simulations, achieving comparable results, which strengthens
greatly our conclusions.

The set up of our runs is highly idealised, since BHs are supposed to be already in
place in a well formed gaseous disc. The latter, at the beginning of the simulation, 
starts cooling and eventually forms stars. Furthermore, our
simulations are evolved in complete isolation, i.e., no gas flows into   
the nuclear disc from larger scales (e.g., from
outer regions of the host galaxy, from cosmological filaments, or
through galaxy mergers). As a consequence, every accretion episode
halts when star formation and SNa--driven gas consumption have
evacuated the central disc regions. 

Though the prescriptions adopted in our simulations regarding star formation and SNa feedback could, in principle, 
affect the growth of BHs, they are conservative for what 
concerns gas accretion onto the BHs. First, the star formation rate in a sphere of radius $\simeq 1$ pc (corresponding to the average clump radius) around BH$_{\rm top}$ is $\simeq 0.1 \rm \msun/yr$, much larger than the average BH accretion rate ($\simeq 10^{-3}\rm \msun/yr$). Hence, we can be confident that in our simulations the gas is mostly consumed by star formation rather than by BH accretion. To further prove the point we run a low resolution \gizmo simulation  in which we increase the star formation efficiency to its maximum value, and find that even for the resulting extremely high star formation rate ($\sim 40\msun/$yr) the BH accretion history is not significantly modified. Second, our assumed timescale for SNa explosions (1 Myr) is  shorter than the typical lifetime of low metallicity stars in the mass range $8-40\msun$ ($\gsim 4$ Myr; \citet{hurley00}). Our resulting SNa feedback is then already very highly efficient\footnote{We run a further \gizmo low resolution simulation and checked that only an unrealistic ``maximally efficient" SNa feedback (in which stars explode as soon as they form) is able to evacuate the gas from the centre before the stellar-size BHs can start to accrete.}.
In this context, we find that SNe produce a high velocity wind ($v_{\rm ej} \lesssim 3000\,\rm km/s$), which can expand up to 5 kpc above the disc plane. In principle such gas could form a galactic fountain falling back on to the disc, allowing for a new phase of super-critical accretion. 

The cooling function employed in our simulations does not take into account molecular hydrogen, which could induce stronger fragmentation in the disc, and enhance SF. This could, in principle, limit the accretion on to the BHs. However, the presence of radiative feedback from stars (not considered in this study) would act in the opposite direction, dissociating molecular hydrogen and thus allowing for higher inflow rates towards the BHs. Both the modelling of the large scale galactic potential (essential to assess the fate of the SNa driven wind) and the effect of molecular hydrogen cooling (and related stellar dissociating flux) are beyond the scope of our study. 

Regardless the spatial/mass resolution and the kind of hydrodynamical
code used, a coherent picture emerges. If BHs have to grow by 2-3 order
of magnitudes in mass, radiative inefficient accretion is a necessary condition, 
but not a sufficient one. 
BHs must find themselves embedded in gas structures that need to be: (i) 
massive enough to provide the gas reservoir, and (ii) dense enough to survive feedback. 
This  may occur when the cooling
gas fragments in clumps, and some of the BHs bind to
them. Such process allows some of the BHs to reach masses as high as $10^3$
- $10^4 \msun$ on Myr timescales, making them viable candidates as 
seeds of the supermassive variety of BHs powering high redshift quasars. 

Mass accretion onto the BHs depends upon the number,
mass and density of the clumps forming in the disc. We showed that these parameters are affected  
by the numerical resolution achieved in the different runs and, as discussed in
Section~\ref{sec:results}, different resolutions result in different
BH accretion histories. We are unable to describe gas dynamics down to the accretion disc scales, even at the highest spatial resolution reached, and this limits our ability to achieve firm estimates of accretion rate and mass growth of the BHs. Yet, the dynamics of gas leading to the formation of dense clouds we observe in all our runs is strongly independent of 
sub--grid recipes. 
The gas within the accretion radius of BHs is far from being rotationally supported. Since the relative gas-BH velocity becomes negligible after the capture process, the gas in fact experiences almost radial inflow toward the BHs. Our estimate of the accretion rate is of the same order of the 
Bondi accretion rate given the temperature and density of the medium surrounding the BHs.

Therefore, despite our accretion histories are not accurate enough from a quantitative point of view, we can be confident about the reliability of the BH-clumps-capture process we observe. Our work should be considered as a {\it proof of concept}, robust enough to
understand under which conditions and through which processes a cluster of stellar mass BHs
can actually experience episodes of super--Eddington growth, and what are the effects 
on the environment.

%While the qualitative behaviour is fairly similar in all runs,  quantitative convergence in terms of accreted mass would require simulations with a mass and spatial
%resolutions even higher than those of high\_G run. Such an extensive
%numerical experiment is out of the scope of the present investigation,
%that, because of the very idealised initial conditions, should be
%considered as a test of the BH-clumps-capture process as trigger of
%super--Eddington accretion events.

We conclude that a radiatively inefficient accretion, together with the aforementioned BH-clumps-capture process, can result in mass growths 10-100 times larger than in the radiatively efficient case, making this mechanism a viable candidate to grow massive BH seeds from stellar mass BHs.

The process we studied can result in a
prolonged super--Eddington accretion phase only as long as the masses of the
clumps are comparable or larger than the masses of the accreting BHs. While the
gravitational capture itself easily binds small clumps to comparatively 
massive BHs, the available gas reservoir is not sufficient
for significant BH growth. Moreover, even feedback from
radiative inefficient accretion severely affects such small
clumps. 

%If accretion continues to be
%radiatively inefficient as the black hole approaches the
%supermassive size,  its effect on the galactic scale
%environment could be in principle weaker than normally assumed in popular AGN feedback
%recipes used in galaxy formation, in which radiation is assumed to percolate efficiently
%to large radii (well above the CND scale). On the other end, given the high
%luminosities reached in the super-critical phases momentum transfer mediated
%by photon pressure, could play an important role if the optical depth is large across the
%nuclear region (Lucio: maybe give an optical depth range and cite the radiation pressure-driven
%wind papers now in fashion for stellar/SN feedback, such as from Faucher-Giguere and Norm
%Murrey?), driving a radiation pressure-driven wind possibly reaching out of the
%CND region. 

%In other words, given the conditions fur super-critical accretion considered here 

Other feedback processes, e.g., momentum-driven feedback, might be important, 
and will be explored in the future. If, however, such processes turn out to be inefficient, this could naturally allow the galaxy to remain highly star--forming despite the fast growth of the MBH, perhaps explaining
the new puzzling observation of a high-z star forming galaxy hosting an SMBH well overweight
for its stellar mass \citep{Trakhtenbrot15}.

In addition, as soon as a BH becomes significantly heavier
than typical gas clumps, it starts migrating toward the centre of the
disc via dynamical friction. This process will naturally bring the
most massive BH (the one that by chance had the largest mass growth, i.e., BH$_{\rm top}$)
to the centre of the host galaxy, where MBHs are commonly observed. At
this point, however, further clumps forming in the disc no longer interact with the central BH.

In order for the large nuclear gas reservoir assumed in our initial conditions to be present in the
galactic nucleus disc angular momentum needs to be removed well 
before gas turns into stars, so that inflowing material can be, at least partially, accreted
by the central BH. This is of course the longstanding fuelling problem of MBH debated in the community \citep[e.g.,] [and references therein]{hicks13}, and its discussion is beyond the goal of the present study. 
Paper II will deal with the nuclear properties and gas inflows in a galaxy at $z \sim 6-10$
which by $z = 3$ will have a mass comparable to that of  the observed clumpy star forming discs, thus allowing to place our model
more properly in the context of galaxy formation and test its assumptions and outcomes. Preliminary analysis shows that the mass
enclosed within a hundred pc scale fluctuates between a few times $10^7$M$_{\odot}$ and just above $10^8$M$_{\odot}$. So, as mentioned in Section 2, our initial conditions seem to be well motivated. 

We finally note that, whenever  inflowing gas refills the circum--nuclear disc, the whole process 
we simulated is rejuvenated: a new massive BH seed will be formed, sinking to the centre of the 
galaxy and eventually forming an intermediate massive black hole binary bound to coalesce owing to gravitational 
radiation losses. This kind of systems may be a perfect target for the planned eLISA observatory \citep{amaroseoane13}.

\section{Acknowledgements}

We thank the anonymous referee for his/her comments and suggestions that helped us to improve the quality of the paper. AL, FH and MD acknowledge financial support from the Italian MIUR, through PRIN 2010-2011. Simulations were run on the EURORA cluster at CINECA and on the Lucia cluster at 
DiSAT, University of Insubria. P.M. acknowledges support by the NSF through grant AST-1229745 and NASA through grant NNX12AF87G.

\bibliographystyle{mn2e}
\bibliography{Slim}
\label{lastpage}

\end{document}